\documentclass[letterpaper,titlepage,11pt]{article}
\usepackage[colorlinks=true,urlcolor=blue,citecolor=magenta]{hyperref}
\usepackage{amssymb,amsmath,amsfonts}
\usepackage{epsfig}
\usepackage{graphicx}
\usepackage{caption}
\usepackage{subcaption}
\usepackage{amscd}
\usepackage{amsthm}
\usepackage{latexsym}
\usepackage{amsbsy}
\usepackage{bbm}
\usepackage[english]{babel}
\usepackage{psfrag}
\usepackage{tabularx}

\def\be{\begin{equation}}
\def\ee{\end{equation}}
\def\bea{\begin{eqnarray}}
\def\eea{\end{eqnarray}}

\allowdisplaybreaks

\def\clock{{\count0=\time
           \divide\count0 60
           \ifnum\count0<10 0\fi\the\count0
           \multiply\count0 -60 \advance\count0 \time
           :\ifnum\count0<10 0\fi \the\count0
         }}
\newcommand{\timestamp}{{\small\vbox{\hbox{\tt\jobname.tex}
\hbox{\the\day/\the\month/\the\year, \clock}}}}

\numberwithin{equation}{section}

\begin{document}

\begin{titlepage}
\rightline{\vbox{   \phantom{ghost} }}
%
%
 \vskip 1.4 cm
\centerline{\LARGE \bf   Symmetries and Couplings of }
\vspace{.2cm}
\centerline{\LARGE \bf  Non-Relativistic Electrodynamics}

\vskip 1.5cm

\centerline{ {\bf  Guido Festuccia$^1$,  Dennis Hansen$^2$, Jelle Hartong$^3$, Niels A. Obers$^2$}}

\vskip .8cm

\begin{center}

\sl $^1$ Department of Physics and Astronomy, Uppsala University, \\ 
\sl  SE-751 08 Uppsala, Sweden \\
\sl $^2$ The Niels Bohr Institute, Copenhagen University,\\
\sl  Blegdamsvej 17, DK-2100 Copenhagen \O , Denmark. \\ 
\sl $^3$ Physique Th\'eorique et Math\'ematique and International Solvay Institutes,\\
Universit\'e Libre de Bruxelles, C.P. 231, 1050 Brussels, Belgium

\vskip 0.4cm

\end{center}
\vskip 0.6cm


\vskip .8cm \centerline{\bf Abstract} \vskip 0.2cm \noindent

We examine three versions of non-relativistic electrodynamics, known as the electric and magnetic limit theories of Maxwell's equations and Galilean electrodynamics (GED) which is the off-shell non-relativistic limit of Maxwell plus a free scalar field. For each of these three cases we study the couplings to non-relativistic dynamical charged matter (point particles and charged complex scalars). The GED theory contains besides the electric and magnetic potentials a so-called mass potential making the mass parameter a local function. The electric and magnetic limit theories can be coupled to twistless torsional Newton--Cartan geometry while GED can be coupled to an arbitrary torsional Newton--Cartan background. The global symmetries of the electric and magnetic limit theories on flat space consist in any dimension of the infinite dimensional Galilean conformal algebra and a $U(1)$ current algebra. 
For the on-shell GED theory this symmetry is reduced but still infinite dimensional, while off-shell only the Galilei algebra plus two dilatations
remain. Hence one can scale time and space independently, allowing Lifshitz scale symmetries for any value of the critical exponent $z$.

\end{titlepage}

\tableofcontents

\section{Introduction}

Maxwell's theory of electromagnetism is one of the cornerstones of modern physics being the first theory that incorporates Lorentz invariance, thus playing a crucial role in the development of special relativity. Nevertheless there are reasons why it is interesting to study non-relativistic limits of the theory, as first considered in the pioneering paper by Le Bellac and L\'evy-Leblond \cite{LeBellac:1973}. As is often the case in physics, by considering limits one may learn more about properties of the theory and in particular in the case of electromagnetism it may teach us which electromagnetic effects are truly relativistic and which ones are not. Moreover, it is interesting to see whether and how precisely one can define a consistent limit of electromagnetism, including Maxwell's field equations and the Lorentz force, and how the corresponding fields transform under Galilean symmetries.  More generally, these theories are non-trivial examples of non-relativistic dynamical theories and from a certain point of view the natural theories to which one may wish to couple charged non-relativistic matter. 

In fact, as also emphasized in \cite{LeBellac:1973}, one may wonder what type of electromagnetism a post-Newtonian but pre-Maxwellian physicist would have written down guided by Galilean invariance. For instance, when one gives up Lorentz symmetry there is going to be a different interplay between symmetries and the continuity equation of charge and current. One may also ask what symmetry structures non-relativistic theories of electromagnetism exhibit and how one can couple these theories to charged point particles and other types of charged matter. Finally, a natural question to ask is how non-relativistic electrodynamics can be covariantly coupled to an appropriate background geometry.

In this paper we will in part revisit these questions and also address a number of new ones, which are especially intriguing in view of the renewed interest in Newton--Cartan (NC) geometry as the non-relativistic background geometry to which one can covariantly couple non-relativistic field theories.%
\footnote{There is a growing literature on this in the last three years. Early papers include \cite{Son:2013rqa} which 
introduced NC geometry to field theory analyses of problems with strongly correlated electrons, such as the fractional quantum Hall effect. 
The novel extension with torsion was first observed as the background boundary geometry in  holography for $z=2$ Lifshitz geometries 
\cite{Christensen:2013lma,Christensen:2013rfa} and a large class of models with arbitrary $z$ \cite{Hartong:2014oma}. 
Further field theory analysis can be found in \cite{Geracie:2014nka,Jensen:2014aia,Hartong:2014pma,Hartong:2015wxa}. Some of the later works that are
relevant in the context of the present paper, dealing with aspects of the coupling to non-relativistic electrodynamics, are  \cite{Duval:2014uoa,Jensen:2014aia,Geracie:2015dea,Bleeken:2015ykr,Bergshoeff:2015sic}.}
In particular, focussing first on flat backgrounds, we will present new angles on the various non-relativistic limits considered in the literature, find 
novel effects and phenomena when coupling non-relativistic electromagnetism to charged particles and matter fields and uncover
new extended symmetries of the theories. 

Moreover, we will show how one can couple non-relativistic electrodynamics to the most general torsional Newton--Cartan (TNC) geometry \cite{Christensen:2013lma,Christensen:2013rfa,Hartong:2014oma} or, as turns out to be relevant in some cases, twistless torsional Newton--Cartan  (TTNC) geometry. This is also interesting in light of our recent work \cite{Festuccia:2016awg} in which (linearized) TNC geometry is shown to arise by applying  the Noether procedure for gauging space-time symmetries  to theories with Galilean symmetries, including both massless and massive realizations. This analysis shows that even in the massless case  it is necessary to introduce the Newton--Cartan one-form $M_\mu$, which 
couples to a  topological current in that case, while for the massive case it couples to the conserved mass current.  Non-relativistic electrodynamics
(in particular Galilean electrodynamics, see below)  is a prominent example of a massless non-relativistic theory. The coupling of
non-relativistic electrodynamics to TNC geometry  derived in this paper provides a nice check with the general linearized results obtained in \cite{Festuccia:2016awg} with the Noether method, including the particular form of the topological current.

Besides the above-mentioned motivations, there are a number of further reasons for our study originating from holography, field theory and gravity.
TNC geometry was first observed  \cite{Christensen:2013lma,Christensen:2013rfa,Hartong:2014oma}  as the boundary geometry in holography for Lifshitz space-times in the bulk (see \cite{Taylor:2015glc} for a review on Lifshitz holography), characterized by anisotropic
scaling between time and space.  If one wishes to consider these systems at finite charge density, e.g. by adding a bulk Maxwell field, one might expect
non-relativistic electromagnetic potentials on a TNC geometry to appear as background sources in the boundary theory.  

Furthermore, it was shown in \cite{Hartong:2015zia} that
dynamical NC geometry correspond to the known versions of (non)-projectable Ho\v rava-Lifshitz (HL) gravity. For these dynamical non-relativistic gravity theories
it is interesting in its own right to examine how they couple to non-relativistic electrodynamics, being the non-relativistic analog of Einstein--Maxwell theory.
This will be moreover relevant for using HL-type gravity theories as bulk theories in holography 
\cite{Griffin:2012qx,Janiszewski:2012nf}. 
In line with this, it was recently shown that three-dimensional HL gravity theories can be written as Chern-
Simons gauge theories on various non-relativistic algebras, including a novel version of non-projectable
conformal Ho\v rava--Lifshitz gravity, also referred to as Chern--Simons Schr{\"o}dinger gravity \cite{Hartong:2016yrf}. These theories are again interesting in holography
but also as effective field theories for condensed matter systems, and one may wonder whether there are likewise Chern--Simons versions
of non-relativistic electromagnetism.   

As a final motivation we note that NC geometry and gravity can be made compatible with supersymmetry \cite{Andringa:2013mma,Bergshoeff:2015uaa,Bergshoeff:2015ija,Knodel:2015byb,Bergshoeff:2016lwr}, 
and thus can provide tools to construct non-relativistic supersymmetric field theories on curved backgrounds, following the relativistic case 
\cite{Festuccia:2011ws}
potentially allowing to employ powerful localization techniques to compute certain observables \cite{Pestun:2007rz}. A particularly interesting case here
could be a quantum mechanically consistent supersymmetric version of non-relativistic electromagnetism, for which our results  could provide
useful input.

An outline and main results of the paper is as follows. 
We start in Section \ref{sec:NRlimit} by reviewing three Galilean invariant non-relativistic theories of electromagnetism in the absence of sources. 
These include the electric theory and magnetic  theory of   \cite{LeBellac:1973} as well as a larger theory \cite{Santos:2004pq}%
\footnote{See also Refs.~\cite{DeMontigny:2005oib,Rousseaux:2013}.},  
which we call Galilean Electromagnetism, and which includes the former two. For GED it is possible to
 find an off-shell formulation, which is not the case for the electric and magnetic theory.  
 Obtaining GED from a non-relativistic limit requires
to add a scalar field to Maxwellian electromagnetism before taking the limit, as described in \cite{Bergshoeff:2015sic}.%
\footnote{See also \cite{Jensen:2014wha} in which non-relativistic limits are revisited.} 
The non-relativistic limits from which these three theories are obtained are discussed, while we also
show how to obtain GED via a null reduction of the Maxwell action in one dimension higher.%
\footnote{For some early work on null reductions see e.g. \cite{Duval:1984cj,Julia:1994bs} and the connection
between null reduction and GED was also discussed in \cite{Duval:1990hj}.}

We then turn in  Section \ref{sec:particles} to the coupling of charged massive point particles in the three different limits of electromagnetism.
Depending on the case, there are a number of interesting features, in terms of the backreaction of the particle on the non-relativistic
electromagnetic fields and the dynamics (forces) that a charged massive particle experiences in a given background. 
 In particular, we will see that for the case of GED the particles act as a source for all gauge invariant fields, and that the force
term includes electric and magnetic components but also a novel contribution. The interpretation of this is that one of the three GED fields 
describes a mass potential, which thus supplements the electric and magnetic fields of the theory. 
We will also show that the minimal coupling of GED to point particles can be obtained by null reduction of the charged point particle in Maxwell theory. 
Section \ref{sec:scalar} treats the electric, magnetic and GED limits for scalar electrodynamics, and we will observe a number of
parallels with the results for charged point particles.  

In Section \ref{sec:symmetries} we study  the symmetries of the three limit theories, by determining the most general set of (linear) transformations
of the fields that leave the theories invariant.  The main result is that the on-shell electric and magnetic theory have a very large invariance group 
containing (in any dimension) both the infinite Galilean conformal algebra and a $U(1)$ current algebra as subgroups. 
Our results are consistent with the results in \cite{Bagchi:2014ysa} for these two theories%
\footnote{Symmetries of non-relativistic electrodynamics were also studied from the Newton-Cartan point of view in \cite{Duval:2009vt}.}, but we find a larger
 symmetry algebra,  as this paper did not consider the most general ansatz.  
We also show that the on-shell GED theory has a smaller set of symmetries, though still infinite dimensional. Furthermore, we show that in the specific
case of   $3+1$ dimensions the finite Galilean conformal algebra is a symmetry. Finally the off-shell GED theory has only the Galilean algebra extended with two dilatations as its invariance group. The two dilatations originate from the fact that we can independently rescale time and space, and
as a consequence we conclude that the GED action has Lifshitz scale invariance for any value of $z$.

The general covariant coupling of the three theories to arbitrary curved non-relativistic spacetime backgrounds, i.e. TNC geometry is presented in Section \ref{sec:TNC}.  After giving a brief review of TNC geometry, we first treat the GED case  which is the simplest case, as it admits a Lagrangian description.
We also show that the resulting action can also be obtained by a null reduction from Maxwellian electromagnetism coupled to a Lorentzian metric.
The linearized version of the GED action coupled to TNC geometry agrees with the one obtained in \cite{Festuccia:2016awg} via
the Noether procedure. 
We then give the covariant form of the equations of motion for the magnetic and electric theories, and in both cases it is found that the spacetime background should be restricted to twistless torsional Newton--Cartan (TTNC) geometry. 
We conclude the section by constructing a covariant minimal coupling to charged scalar fields, which can be obtained as well from null reduction
of scalar QED in one dimension higher coupled to a Lorentzian metric, and generalize this to non-minimal couplings. 
We end the paper in Section \ref{sec:outlook} with some interesting open problems.

\section{Non-relativistic limits of Maxwell's equations}\label{sec:NRlimit}

In this section we will discuss how to obtain Galilean invariant theories by taking a non-relativistic limit of electromagnetism.
Following the seminal work \cite{LeBellac:1973} there are two such limits usually referred to as the ``electric" limit and the ``magnetic" limit. We will review how these limits arise and show how they can both be embedded in a larger theory \cite{Santos:2004pq}  which we will refer to as Galilean Electromagnetism (GED). For simplicity in this section we will work in the absence of sources which will be added later. 

Consider a $U(1)$ gauge field $A_\mu$ in Minkowski space-time with Cartesian coordinates $(t, x_i)$. The gauge transformations are given by
$$
A'_t= A_t+{1\over c} \partial_t \Lambda~,\qquad A'_i=A_i+\partial_i \Lambda
$$
while the equations of motion  $\partial_\mu F^{\mu\nu}=0$ read explicitly: 
\begin{equation}\label{eq:Maxwell}
\partial_i\left(\partial_i A_t-\frac{1}{c}\partial_t A_i\right)=0\,,\qquad \frac{1}{c}\partial_t\left(\partial_i A_t-\frac{1}{c}\partial_t A_i\right)+\partial_jF_{ji}=0~.
\end{equation}
Here $c$ is the speed of light and  $F_{ij}=\partial_i A_j-\partial_j A_i$. There exist two non-relativistic limits known as the electric and magnetic limits, depending on whether the vector potential $A_\mu$ is timelike or spacelike, respectively. 

The  ``electric" limit of these equations can be obtained as follows
\begin{equation}
\text{(electric limit)}\qquad A_t=-\varphi\,,\qquad A_i={1\over c} a_i\,,\qquad\Lambda={1\over c}\lambda~,\qquad \text{$c\rightarrow\infty$ with $\varphi$, $a_i$, $\lambda$ fixed.} 
\end{equation}
This results in 
\begin{equation}\label{eq:Elimit}
\partial^i\partial_i\varphi=0\,,\qquad -\partial_t\partial_i\varphi+\partial^jf_{ji}=0\,,
\end{equation}
where $f_{ij}=\partial_i a_j-\partial_j a_i$ is the magnetic field. The contraction of the relativistic gauge transformations leads to
$\delta_\lambda\varphi=0$ and $\delta_\lambda a_i=\partial_i\lambda$ so that the scalar $\varphi$ is invariant.
The equations \eqref{eq:Elimit} respect a symmetry under Galilean boosts $x'^i=x^i+v^i t$, $t'=t$ acting on the the fields $\varphi$ and $a_i$ as  $$\varphi'=\varphi~\qquad a'_i=a_i+v_i\varphi~.$$ This follows from first performing a Lorentz boost on $A_\mu$ and then taking the $c\rightarrow\infty$ limit. 

The  ``magnetic" limit  can be similarly defined by setting
\begin{equation}
\text{(magnetic limit)}\qquad A_t=-\tilde\varphi\,,\qquad A_i= c a_i\,,\qquad\Lambda= c \lambda~,\qquad  \text{$c\rightarrow\infty$ with $\tilde\varphi$, $a_i$, $\lambda$ fixed.}
\end{equation}
In this case the equations of motion  \eqref{eq:Elimit} reduce to
\begin{equation}\label{eq:Mlimit}
\partial_i\tilde E^i=0\,,\qquad \partial^jf_{ji}=0\,,
\end{equation}
where $\tilde E_i=-\partial_i\tilde\varphi-\partial_ta_i$ is the electric field. Gauge transformations act as $\delta_\lambda\tilde \varphi=-\partial_t \lambda$ and $\delta_\lambda a_i=\partial_i\lambda$ so that the electric field is invariant. In this limit the potentials $\tilde\varphi$ and $a_i$ transform under Galilean boosts as 
\begin{equation}
\label{eq:magtra}
\tilde\varphi'=\tilde\varphi+v^i a_i~,\qquad a_i'=a_i~.
\end{equation}
In 3+1 dimensions the electric and magnetic limits are related by electric/magnetic duality \cite{Bleeken:2015ykr}.

Finally we can define a third limit that has the advantage of allowing an off-shell description. Consider the Maxwell action for $A_\mu$ with an additional free real scalar field $\chi~,$
\begin{equation}\label{eq:Maxwellscalar}
\mathcal{L}=\frac{1}{2 c^2}\left(\partial_t A^i-c\partial^i A_t\right)\left(\partial_t A_i-c\partial_i A_t\right)-\frac{1}{4}F^{ij}F_{ij}+\frac{1}{2c^2}\partial_t\chi\partial_t\chi-\frac{1}{2}\partial^i\chi\partial_i\chi\,.
\end{equation}
The limit is given by
\begin{equation}\label{eq:GEDlimit}
\text{(GED limit)}\qquad \chi = c\varphi\,,\qquad A_t = -c\varphi-\frac{1}{c}\tilde\varphi\,,\qquad A_i=a_i\,,\qquad\text{$c\rightarrow\infty$ with $\varphi$, $\tilde\varphi$, $a_i$ fixed.}
\end{equation}
By substitution in \eqref{eq:Maxwellscalar} we obtain the action for Galilean electrodynamics (GED)
\begin{equation}\label{eq:flatGED}
S=\int d^{d+1}x\left(-\frac{1}{4}f^{ij}f_{ij}-\tilde E^i\partial_i\varphi+\frac{1}{2}\left(\partial_t\varphi\right)^2\right)\,.
\end{equation}
This action was first introduced in \cite{Santos:2004pq} and the limit from which it arises is described in \cite{Bergshoeff:2015sic}~. Under gauge transformations the fields transform as $$\delta_\Lambda\tilde\varphi=-\partial_t\Lambda~, \quad \delta_\Lambda a_i=\partial_i\Lambda~,\quad\delta \varphi=0~.$$
The action \eqref{eq:flatGED} is invariant under Galilean boosts acting on the fields as 
\begin{equation}
\label{eq:GEDboosts}
\tilde\varphi'=\tilde\varphi+v^i a_i+\frac{1}{2}v^i v^i\varphi~,\quad a'_i=a_i+v_i\varphi~,\quad \varphi'=\varphi~.
\end{equation}
The equations of motion are given by \eqref{eq:Elimit} together with an additional equation of motion obtained by varying $\varphi$ which reads
\begin{equation}\label{eq:extraeom}
\partial_t^2\varphi-\partial_i\tilde E_i=0\,.
\end{equation}

At this point it could be argued that the action \eqref{eq:flatGED} provides an off-shell formulation of the electric limit because its equations of motion comprise \eqref{eq:Elimit} and \eqref{eq:extraeom} does not further constrain $a_i$ nor $\varphi$ and can be used to solve for $\tilde \varphi$. 
There are however a number of reasons why these should be considered as separate theories.
\begin{itemize}
\item{In section \ref{sec:symmetries} we will show that the symmetries of \eqref{eq:Elimit} comprise a larger set of transformations than the symmetries that preserve the GED equations of motion.}
\item{As will see in the next section the two theories couple to sources with distinct properties.}
\item{In sections \ref{subsec:magTNC} we will show that the two theories couple differently to curved space.}
\end{itemize}

The magnetic limit \eqref{eq:Mlimit} arises from the equations of motion of GED by noticing that it is consistent to set $\varphi=0$ in \eqref{eq:Elimit} and \eqref{eq:extraeom}. We are not aware of an action for the magnetic limit fields $\tilde\varphi$ and $a_i$ (and potentially other fields) whose equations of motion lead to \eqref{eq:Mlimit}.

\subsubsection*{GED from null reduction}
Another way to obtain the GED action is by performing a null reduction of the Maxwell action in one higher dimension. Indeed consider the $d+2$ dimensional Maxwell action
\begin{equation}
S=\int dudtd^dx\left(-\frac{1}{4}\eta^{AC}\eta^{BD}F_{AB}F_{CD}\right)\,,
\end{equation}
where $\eta_{AB}dX^A dX^B=2dtdu+dx^i dx^i$. We can now set $A_u=\varphi$, $A_t=-\tilde\varphi$ and $A_i=a_i$ and impose that all the fields are independent of the $u$ coordinate to obtain the GED action \eqref{eq:flatGED}. We will generalize this null reduction to the case of a curved background in section \ref{subsec:nullredcr}.

The three limits discussed here can be expressed in terms of three different gauge invariant quantities: the electric field $\tilde E_i$, the magnetic field $f_{ij}$ and the scalar $\varphi$. We will show that $\varphi$ should not be interpreted as an electric potential. Instead we will refer to $\varphi$ as a mass potential for reasons that will become clear in the next section as well as in section \ref{sec:matter}. 

\section{Coupling to point particles}\label{sec:particles}

Here we will consider how to couple the different limits of electromagnetism we discussed in the last section to charged massive particles. As $c\rightarrow \infty$ the particles are slowly moving (non-relativistic). We will see that in the electric limit the point charges experience only electric forces but act as a source both for the electric and magnetic fields. In the magnetic limit the Lorentz force is also Galilean invariant but the charged particles do not backreact on the magnetic fields. Finally for GED the particles act as a source for all gauge invariant fields. In this case the forces acting on the charged particles are both of electric and magnetic form but also of a novel kind for which we will put forward an interpretation.

The Lagrangian density for a relativistic point particle of mass $m$ and charge $q$ minimally coupled to the Maxwell gauge potential is given by
\begin{equation}\label{eq:pointsourceaction}
\mathcal{L}=\left(-mc^2 \sqrt{1-\frac{\dot X^i\dot X^i}{c^2}}+mc^2+qA_t+qA_i\frac{\dot X^i}{c}\right)\delta^{(d)}(\vec x-\vec X(t))\,,
\end{equation}
where $\vec X(t)$ is the position of the particle at time $t$.   

We can add  \eqref{eq:pointsourceaction} to the lagrangian for the gauge fields given by \eqref{eq:Maxwellscalar} (excluding the uncoupled scalar field $\chi$). This results in the following equations of motion for the gauge fields and $X(t)$,
\begin{eqnarray}
&&\partial_i\left(\partial_i A_t-\frac{1}{c}\partial_t A_i\right)=q\delta^{(d)}(\vec x-\vec X(t))\,,\nonumber\\
&& \frac{1}{c}\partial_t\left(\partial_i A_t-\frac{1}{c}\partial_t A_i\right)+\partial_jF_{ji}=-q\frac{\dot X^i}{c}\delta^{(d)}(\vec x-\vec X(t))\,,\cr
&&m \frac{d}{dt}\left(\frac{\dot X^i}{\sqrt{1-\frac{\dot X^j\dot X^j}{c^2}}}\right)=q\left(\partial_i A_t-\frac{1}{c}\partial_t A_i\right)+\frac{q}{c}\dot X^j F_{i j}\,.
\end{eqnarray}
In the electric limit these equations reduce to:
\begin{eqnarray}
\label{eq:pointelec}
&&\partial_i \partial_i \varphi=- q\delta^{(d)}(\vec x-\vec X(t))\,,\cr
&& \partial_t \partial_i \varphi-\partial_jf_{ji}=q{\dot X}^i\delta^{(d)}(\vec x-\vec X(t))\,,\cr
&&m{\ddot X^i}=-q \partial_i \varphi\,.
\label{eq:elimsource}
\end{eqnarray}
Hence the charged particle sources both the magnetic field $f_{ij}$ and electrostatic potential $\varphi$ but is not acted upon by the magnetic field. This is consistent with the analysis presented in \cite{LeBellac:1973} where it was found that slowly moving charges generate fields of the electric kind and that in this limit it is only possible to describe electric forces (whence the name).

In the magnetic limit we obtain instead:
\begin{eqnarray}\label{eq:pointsourcedmagnetic}
&&\partial_i\tilde E_i=q\delta^{(d)}(\vec x-\vec X(t))\,,\qquad \partial_jf_{ji}=0\,.\\
&&m{\ddot X^i}=q \tilde E^i+q \dot X^j f_{ij}\,.\nonumber
\end{eqnarray}
In this case the particle is acted upon by both electric and magnetic forces but does not source the magnetic field which can be considered as an external background. Because the particle is slowly moving this procedure does not give rise to the most general source terms that can be consistently coupled to the electromagnetic fields in the magnetic limit. Indeed according to \cite{LeBellac:1973} it is possible to introduce a charge density $\rho(x)$ and a current $J_i(x)$ whose divergence vanishes $\partial_i J^i=0$ and is not related to charge transport (so that there is no continuity equation relating $\rho$ and $J_i$). These can then act as sources for the electric and magnetic fields 
\begin{equation}\label{eq:generalsourcedmagnetic}
\partial_i\tilde E_i=\rho\,,\qquad \partial_jf_{ji}=J_i~.
\end{equation}
In order to ensure invariance under Galilean boosts the sources\footnote{Sources $J_i$ and $\rho$ with these properties can be constructed starting with configurations of charges in the relativistic theory such that $J_i \sim c \rho$ and taking the magnetic limit.}  have to transform as $J'=J$ and $\rho'=\rho+ v^i J_i$. As a consequence the only force term involving these sources that stays the same in different inertial reference frames is of magnetic type
$$
 F_i= \int d^3 x f_{i j} J^j~.
$$

Next we will couple charged particles to GED. We will consider the Lagrangian  \eqref{eq:Maxwellscalar} for the Maxwell gauge fields coupled to the scalar $\chi$ and add to it  the Lagrangian density for the point particle \eqref{eq:pointsourceaction}. In order to obtain a finite non-relativistic limit we will also introduce a coupling between $\chi$ and the point particle whose form is reminiscent of the dilaton coupling  to a D-brane Nambu-Goto action,
$$
\Delta \mathcal{L}_{GED}=q \chi \sqrt{1-\frac{\dot X^i\dot X^i}{c^2}}\delta^{(d)}(\vec x-\vec X(t)) ~.
$$
We can then take the limit $c\rightarrow \infty$ keeping $\hat q=  q/c $ constant while the Maxwell fields and $\chi$ scale as in \eqref{eq:GEDlimit}. As a result the GED fields described by \eqref{eq:flatGED} couple to the point charge according to
\begin{equation}\label{eq:pointGED}
\mathcal{L}=\left(\frac{1}{2}\left(m-\hat q\varphi\right)\dot X^i\dot X^i-\hat q\tilde\varphi+\hat qa_i\dot X^i\right)\delta^{(d)}(\vec x-\vec X(t))\,.
\end{equation}
Hence the equations of motion for the GED fields \eqref{eq:Elimit} and \eqref{eq:extraeom} are modified to:
\begin{eqnarray}
&&\partial_i\partial_i\varphi=-\hat q \delta^{(d)}(\vec x-\vec X(t))\,,\label{eq:GEDlimitpoint1}\\
&&\partial_t\partial_i\varphi-\partial_j f_{ji}=\hat q \dot X^i \delta^{(d)}(\vec x-\vec X(t))~,\\
&&\partial_t^2\varphi-\partial_i\tilde E_i=-\frac{1}{2}\hat q\dot X^i\dot X^i\delta^{(d)}(\vec x-\vec X(t))\,,
\label{eq:GEDlimitpoint2}
\end{eqnarray}
while the equation of motion for the point particle is given by
\begin{equation}\label{eq:pointdynamics}
\frac{d}{dt}\left[\left(m-\hat q\varphi\right)\dot X^i\right]=\hat q\tilde E_i+\hat q\dot X^j f_{ij}-\frac{\hat q}{2}\dot X^j\dot X^j\partial_i\varphi\,.
\end{equation}
It can be checked that these equations of motion are invariant under Galilean boosts acting on the GED fields according to \eqref{eq:GEDboosts}. We see that $\tilde E_i$ acts on the point particle as an electric field and $f_{ij}$ as a magnetic field. The field $\varphi$ couples to the time component of the mass current of the point particle, it effectively changes $m$  to a local function  $m-\hat q\varphi$. We will refer to $\varphi$ as the mass potential. The term $\partial_t^2\varphi$ in equation \eqref{eq:GEDlimitpoint2} has no counterpart in electrodynamics and we cannot remove it by setting $\varphi=0$ consistently. Hence GED coupled to point particles is markedly different from what we obtained either for the electric or magnetic limit in  equations \eqref{eq:elimsource} and \eqref{eq:pointsourcedmagnetic}. Nevertheless GED also arises from the non-relativistic limit of a relativistic theory albeit one that contains a real scalar field in addition to the gauge fields. 
Note that because $\varphi(x)$ is boost and gauge invariant there are many non-minimal couplings in addition to those appearing in \eqref{eq:pointGED}. For instance we could add a further term linear in the GED fields
\begin{equation}\label{eq:Coulombterm}
\Delta {\mathcal L}= -{\gamma\over 2} \hat q \,\varphi \,\delta^{(d)}(\vec x-\vec X(t))
\end{equation}

Because we have taken a limit where the speed of light is infinite the GED fields propagate instantaneously. It is therefore easy to determine their values at a given time knowing the distribution of charges (and its time derivatives) at the same time. These fields can then be substituted back in 
\eqref{eq:pointGED} resulting in the following Lagrangian for a collection of point charges $q_i$ with masses $m_i$
\begin{equation}
\label{eq:nonrelDarwin}
L=\sum_i {1\over 2} m_i v_i^2-\gamma \sum_{i\neq j} {q_i q_j \over 4 \pi r_{i j}}-\sum_{i\neq j} {q_i q_j \over 4 \pi r_{i j}}  (v_i-v_j)^2\,. 
\end{equation}
Here $(v_i-v_j)$ is the relative speed between two particles and $r_{i j}$ is their separation.
This is similar in spirit to Darwin's Lagrangian \cite{jackson_classical_1999} describing interactions among pointlike charges in electrodynamics up to order $c^{-2}$ in a large $c$ expansion\footnote{A more apt parallel should perhaps be drawn with Weber's electrodynamics which is also manifestly Galilean invariant.}. However  \eqref{eq:nonrelDarwin} does not involve any approximation. Note that the strength of the Coulomb interaction is set by the arbitrary parameter $\gamma$ appearing in \eqref{eq:Coulombterm}. This is possible because the Coulomb interaction is Galilean invariant by itself. In Darwin's Lagrangian instead the Coulomb term is related to other terms of order ${v^2\over c^2}$ by Lorentz transformations.

\subsubsection*{Minimal coupling from null reduction}

Another way of obtaining the minimal coupling of GED to point particles is by null reduction of Maxwell's theory coupled to a point particle in one dimension higher. At the end of the previous section we already showed that the GED action can be obtained by null reduction of Maxwell's theory. Here we will show that the point particle action obtained from \eqref{eq:pointGED} can be obtained by the null reduction of the action of a massless charged relativistic particle on Minkowski space-time i.e.
\begin{equation}
S=\int d\lambda\left(\frac{1}{2e}\eta_{AB}\dot X^A\dot X^B+\hat qA_A\dot X^A\right)\,.
\end{equation}
Let us take for $\eta_{AB}dX^A dX^B=2dtdu+dX^i dX^i$ so that we find
\begin{equation}
S=\int d\lambda\left(\frac{1}{e}\left(\frac{1}{2}\dot X^i\dot X^i+\dot t\dot u\right)+\hat q\varphi\dot u-\hat q\tilde\varphi\dot t+\hat qa_i\dot X^i\right)\,,
\end{equation}
where we wrote $A_u=\varphi$, $A_t=-\tilde\varphi$ and $A_i=a_i$. We will set the momentum conjugate to $\dot u$ equal to a constant $m$:
\begin{equation}
\frac{\partial L}{\partial\dot u}=\frac{\dot t}{e}+\hat q\varphi=m\,,
\end{equation}
from which we can solve for $e$ and substitute into the action to obtain
\begin{equation}
\label{eq:sged}
S=\int d\lambda\left(\frac{1}{2}(m-\hat q\varphi)\frac{\dot X^i\dot X^i}{\dot t}-\hat q\tilde\varphi\dot t+\hat q a_i\dot X^i\right)\,.
\end{equation}
This action has worldline reparametrization invariance $\delta\lambda=\xi(\lambda)$, $\delta X^\mu=\xi\dot X^\mu$. We can gauge fix this symmetry by setting $\dot t=1$ so that worldline time and coordinate time are the same; this choice reproduces \eqref{eq:pointGED}.

\section{Non-relativistic limits of scalar Electrodynamics}\label{sec:scalar}

In this section we will consider the electric, magnetic and GED limits for scalar electrodynamics drawing parallels with the results of the previous section. The starting point is a massive charged complex scalar minimally coupled to $U(1)$ gauge fields
\begin{equation}\label{eq:scalarQEDrel}
{\cal L}={1\over c^2}(\partial_t -i q A_t )\phi (\partial_t +i q A_t )\phi^\star-\Big(\partial_i -i {q\over c} A_i \Big)\phi \Big(\partial_i +i {q\over c} A_i \Big)\phi^\star- {m^2 c^2}\phi \phi^\star\,,
\end{equation}
giving rise together with  \eqref{eq:Maxwellscalar} to the following equations of motion
\begin{eqnarray}\label{eq:eomsqed}
&&\partial_i\left(\partial_i A_t-\frac{1}{c}\partial_t A_i\right)=i {q\over c^2} (\phi^\star (\partial_t -i q A_t )\phi-\phi (\partial_t +i q A_t )\phi^\star) \,,\nonumber\\
&& \frac{1}{c}\partial_t\left(\partial_i A_t-\frac{1}{c}\partial_t A_i\right)+\partial_jF_{ji}=i{q\over c}\left(\phi^\star\Big(\partial_i -i {q\over c} A_i \Big)\phi- \phi \Big(\partial_i +i {q\over c} A_i \Big)\phi^\star \right) \,,\cr
&&{1\over c^2}(\partial_t -i q A_t )(\partial_t -i q A_t )\phi-\Big(\partial_i -i {q\over c} A_i \Big)\Big(\partial_i -i {q\over c} A_i \Big)\phi+m^2 c^2 \phi=0 \,.
\end{eqnarray}

In order to analyse their various limits we need to specify how to scale the complex scalar fields as $c\rightarrow \infty$. We will define a field $\psi(x,t)$ so that
\begin{equation}
\phi(t,x)={1\over \sqrt{2 m}}e^{-i m c^2 t} \psi (t,x)~. 
\end{equation}
This allows to take a finite limit of the equations of motion for $\phi(x,t)$ where the classical mass $m$ and $\psi(x,t)$ are kept fixed as $c\rightarrow \infty$.

In conjunction with the Electric limit scaling for the gauge fields \eqref{eq:Elimit} the equations of motion \eqref{eq:eomsqed} become
\begin{eqnarray}\label{eq:Elimsqed}
&&\partial_i \partial_i \varphi= - q\, \psi^\star\psi \,,\nonumber\\
&& \partial_t \partial_i \varphi-\partial_j f_{ji}=-i{q\over 2m} \left(\psi^\star \partial_i \psi- \psi \partial_i \psi^\star \right) \,,\cr
&&(\partial_t +i q \varphi )\psi={i\over 2m}\partial_i\partial_i \psi \,.
\end{eqnarray}
Note that in the electric limit the Schr\"odinger field $\psi(t,x)$ is inert under gauge transformations. The last equation of motion is the Schr\"odinger equation coupled to the electrostatic potential $\varphi$. The magnetic fields do not appear in the equation of motion for $\psi$ but this field acts as a source for both the electrostatic potential and the magnetic fields. This is indeed consistent with what we found for pointlike charged particles in the electric limit \eqref{eq:pointelec}. 

As for the magnetic limit described in \eqref{eq:Mlimit} it results in
\begin{eqnarray}\label{eq:Mlimsqed}
&&\partial_i  \tilde E^i= - q\, \psi^\star\psi\,,\qquad \partial_j f_{ji}=0\,,\nonumber\\
&&(\partial_t +i q \tilde \varphi )\psi={i\over 2m}(\partial_i-i q a_i)( \partial_i -i q a_i)\psi \,.
\end{eqnarray}
In this case the Schr\"odinger field varies under gauge transformations and its equation of motion involves couplings to both electric and magnetic fields. However $\psi(t,x)$ sources only electric fields. This is consistent with the point particle case \eqref{eq:pointsourcedmagnetic} as expected.   
Indeed it was recognized in  \cite{Brown1999} that the Schr\"odinger field cannot be coupled to either the electric limit or the magnetic limit of the Maxwell equations in such a way that \footnote {It was argued in \cite{Goldin2001321} that these issues could be overcome by introducing appropriate nonlinearities in the constitutive relations entering the Maxwell equations.}
\begin{itemize}
\item{ The resulting model is Galilean invariant,}
\item{The field $\psi$ sources both electric and magnetic fields,}
\item{Both magnetic and electric couplings to $\psi$ are present.}
\end{itemize}

Next we will move to the coupling to GED. In analogy with the case of point-particles described in the previous section, before taking any limit, we will add to scalar electrodynamics a coupling to the scalar field $\chi$ appearing in \eqref{eq:Maxwellscalar}
\begin{equation}
\Delta {\cal L}={1\over c^2}(2 q m c^2 \chi -q^2\chi^2) \phi\phi^\star\,.
\end{equation}
By sending $c\rightarrow \infty$ keeping $\hat q= q/c$ and $\psi(x,t)$ fixed and with the GED fields scaling as in \eqref{eq:GEDlimit} we get a Lagrangian describing the coupling of the Schr\"odinger model to GED.
\begin{equation}\label{eq:scalarGED}
{\mathcal L}=i{(m-\hat q \varphi)\over 2m}\Big(\psi^\star (\partial_t +i q \tilde \varphi)\psi-\psi (\partial_t -i q \tilde \varphi)\psi^\star\Big)-{1\over 2m} (\partial_i -i q a_i )\psi (\partial_i +i q a_i)\psi^\star\,,
\end{equation}
Also in this case as for the case of point-particles the GED field $\varphi(x)$ plays the role of an effective mass. We are allowed to add non-minimal interactions to \eqref{eq:scalarGED}. For instance, in analogy with \eqref{eq:Coulombterm} we can consider a coupling proportional to $\varphi \psi \psi^\star$.

\section{Symmetries}\label{sec:symmetries}

Here we identify what symmetries are present in the various limits discussed in section \ref{sec:NRlimit}. We will first compute the invariance group of the electric and magnetic limit, i.e. equations \eqref{eq:Elimit} and \eqref{eq:Mlimit}. Then we will work out the invariance group of the on-shell GED theory, i.e. the equations of motion of \eqref{eq:flatGED} which are \eqref{eq:Elimit} and \eqref{eq:extraeom}. Finally we will check which of these on-shell symmetries are symmetries of the action \eqref{eq:flatGED}. We will always assume  that $d>1$. 

The main results are that the on-shell electric and magnetic theory have a very large invariance group that in any dimension contains both the infinite Galilean conformal algebra and a $U(1)$ current algebra as subgroups. The infinite dimensionality comes from the fact that the equations of motion are time reparametrization invariant and from the fact that we can perform time dependent translations as well as time dependent spatial dilatations. The GED theory on-shell has a smaller set of symmetries that is still infinite dimensional due to the freedom to perform time-dependent translations. Here we see that $3+1$ dimensions is special in that this is the only dimension in which the finite Galilean conformal algebra is a symmetry of on-shell GED. Finally the off-shell GED theory has only the Galilean algebra extended with two dilatations as its invariance group. The two dilatations originate from the fact that we can independently rescale time and space. Another way of saying this is that the GED Lagrangian has Lifshitz scale invariance for any value of $z$.

In order to find the most general set of transformations that leave the various theories we described in section \ref{sec:NRlimit} invariant we start by writing down the most general set of linear transformations of all the fields
\begin{eqnarray}
\delta\varphi & = & \xi^t\partial_t\varphi+\xi^k\partial_k\varphi+\alpha_1\varphi+\alpha_2\tilde\varphi+\alpha_1^k a_k\,,\label{eq:sym1}\\
\delta\tilde\varphi & = & \xi^t\partial_t\tilde\varphi+\xi^k\partial_k\tilde\varphi+\alpha_3\varphi+\alpha_4\tilde\varphi+\alpha_2^k a_k\,,\label{eq:sym2}\\
\delta a_i & = & \xi^t\partial_t a_i+\xi^k\partial_k a_i+a_k\partial_i\xi^k+\alpha_3^i\varphi+\alpha_4^i\tilde\varphi+\alpha^{ik} a_k\,,\label{eq:sym3}
\end{eqnarray}
where $\xi^t$, $\xi^k$, $\alpha_1$, etc. are all functions of $t,x^i$. These transformations are written with the understanding that in the case of the electric limit we do not transform any field into $\tilde\varphi$ and likewise in the magnetic limit we do not transform the fields into $\varphi$. 

\subsubsection*{Electric limit}
Demanding invariance of the first equation in \eqref{eq:Elimit}, i.e. that $\partial_i\partial_i\delta\varphi=0$ upon use of the equation of motion, leads to the following conditions
\begin{eqnarray}
&&\alpha_2=0\,,\qquad\alpha_1^k=0\,,\qquad\partial_i\xi^t=0\,,\qquad\partial_i\xi^j+\partial_j\xi^i=2\Omega\delta^{ij}\,,\nonumber\\
&&\partial_i\partial_i\xi^k+2\partial_k\alpha_1=0\,,\qquad\partial_i\partial_i\alpha_1=0\,,
\end{eqnarray}
where $\Omega$ is a function of $t$ and $x^i$. Using these results we find after performing the same analysis for the second equation \eqref{eq:Elimit} the following conditions
\begin{eqnarray}
&&\alpha_1=(d-2)\Omega+\gamma\,,\qquad\alpha_3^k=-\partial_t\xi^k\,,\qquad\alpha_4^k=0\,,\qquad\alpha^{ik}=\left(\alpha_1-2\Omega+\partial_t\xi^t\right)\delta^{ik}\,,\nonumber\\
&&\partial_i\Omega=0\,,\qquad\xi^i=\zeta^i(t)+\lambda^i{}_j x^j+\Omega(t) x^i\,,\label{eq:conditions2}
\end{eqnarray}
where $\gamma$ and $\lambda^i{}_j=-\lambda^j{}_i$ are constants. There are two arbitrary scalar functions of time, namely $\xi^t$ and $\Omega$ and there is one vector $\zeta^i$ whose time dependence is arbitrary. These correspond to time reparametrization invariance ($\xi^t$), time dependent spatial dilatations ($\Omega$) and time dependent spatial translations ($\zeta^i$). The fact that one cannot have time dependent rotations was also observed in \cite{Bagchi:2014ysa}. The symmetries of the electric limit thus constitute a very large group that acts on $\varphi$ and $a_i$ as
\begin{eqnarray}
\delta\varphi & = & \xi^t(t)\partial_t\varphi+\xi^k\partial_k\varphi+\left((d-2)\Omega(t)+\gamma\right)\varphi\,,\nonumber\\
\delta a_i & = & \xi^t(t)\partial_t a_i+\xi^k\partial_k a_i+a_k\partial_i\xi^k-\varphi\partial_t\xi^i+\left((d-4)\Omega(t)+\partial_t\xi^t(t)+\gamma\right) a_i \label{eq:ElimitSym}\\
& = & \xi^t(t)\partial_t a_i+\xi^k\partial_k a_i-\lambda_i{}^k a_k-\varphi\left(\partial_t \zeta^i+x^i\partial_t\Omega\right)+\left((d-3)\Omega(t)+\partial_t\xi^t(t)+\gamma\right) a_i\,,\nonumber
\end{eqnarray}
where $\xi^i$ is given by the expression appearing in \eqref{eq:conditions2}. 

\subsubsection*{Magnetic limit}

If we perform a similar analysis in the case of the magnetic limit we ask for the invariance group of the equations \eqref{eq:Mlimit}. We start with the first of these two equations and demand that we find zero when transforming $\tilde\varphi$ and $a_i$ as in \eqref{eq:sym2} and \eqref{eq:sym3} (with no terms going into $\varphi$).  The transformation of the equation of motion leads to terms that involve two, one and zero derivatives on $\tilde\varphi$ and $a_i$. At each order in derivatives we should demand invariance. Doing this first at 2nd order in derivatives up to the use of the equations \eqref{eq:Mlimit} and then at first order etc we find 
\begin{eqnarray}
&&\partial_i\xi^t=0\,,\qquad\alpha_4^i=0\,,\qquad\alpha_2^i=-\partial_t\xi^i\,,\qquad\partial_i\xi^j+\partial_j\xi^i=2\Omega\delta^{ij}\,,\nonumber\\
&&\alpha^{ij}=\bar\gamma\delta^{ij}\,,\qquad\partial_t\xi^t+\bar\gamma=\alpha_4\,,\qquad\partial_i\partial_i\xi^j=0\,,\qquad\partial_t\left(\partial_i\xi^j-\partial_j\xi^i\right)=0\,,
\end{eqnarray}
where $\bar\gamma$ is a constant. Using these results and repeating the procedure for the invariance of the second equation of \eqref{eq:Mlimit} we obtain the extra condition 
\begin{equation}
\partial_i\left(\partial_i\xi^j-\partial_j\xi^i\right)=0\,.
\end{equation}
From all of the above we derive that $\Omega=\Omega(t)$ and that $\xi^\mu$ takes the same general form as in the case of the electric limit, namely $\xi^t=\xi^t(t)$ and $\xi^i=\zeta^i(t)+\lambda^i{}_j x^j+\Omega(t) x^i$. The difference between the two cases lies in the way in which the fields transform into each other. For the magnetic limit theory the symmetries are
\begin{eqnarray}
\delta\tilde\varphi & = & \xi^t(t)\partial_t\tilde\varphi+\xi^k\partial_k\tilde\varphi-\partial_t\xi^k a_k+\left(\partial_t\xi^t+\bar\gamma\right)\tilde\varphi\,,\nonumber\\
\delta a_i & = & \xi^t(t)\partial_t a_i+\xi^k\partial_k a_i+a_k\partial_i\xi^k+\bar\gamma a_i\,.\label{eq:MlimitSym}
\end{eqnarray}

\subsubsection*{Symmetry generators}

The generator $\xi$ can be written as
\begin{equation}
\xi=\xi^\mu\partial_\mu=\xi^t\partial_t+\zeta^i\partial_i+\left(\partial_t\xi^t+\bar\Omega\right)x^i\partial_i+\lambda^i{}_jx^j\partial_i\,,
\end{equation}
where we defined
\begin{equation}\label{eq:defOmega}
\Omega=\partial_t\xi^t+\bar\Omega\,.
\end{equation}
If we take $t$ to be a complex variable we can perform a Laurent expansion of the functions $\xi^t$, $\zeta^i$ and $\bar\Omega$ as follows
\begin{equation}
\xi^t=-\sum_n a_n t^{n+1}\,,\qquad \zeta^i=\sum_n b_n^i t^{n+1}\,,\qquad \bar\Omega=\sum_n c_n t^n\,.
\end{equation}
Defining
\begin{equation}
\xi=\sum_n\left(a_n L^{(n)}+b_n^iM_i^{(n)}+c_n K^{(n)}\right)-\frac{1}{2}\lambda^{ij}J_{ij}\,,
\end{equation}
this gives rise to the following set of generators
\begin{equation}
L^{(n)} = -t^{n+1}\partial_t-(n+1)t^n x^i\partial_i\,,\qquad K^{(n)} = t^n x^i\partial_i\,,\qquad M^{(n)}_i = t^{n+1}\partial_i\,,
\end{equation}
where $n\in\mathbb{Z}$. These generators satisfy the algebra
\begin{eqnarray}
&&[L^{(n)}\,,L^{(m)}]=(n-m)L^{(n+m)}\,,\qquad [L^{(n)}\,,M_i^{(m)}]=(n-m)M_i^{(n+m)}\,,\nonumber\\
&&[K^{(n)}\,,M_i^{(m)}]=-M_i^{(n+m)}\,,\qquad [L^{(n)}\,,K^{(m)}]=-mK^{(n+m)}\,,
\end{eqnarray}
with all other commutators zero. The rotation generators commute with $L^{(n)}$ and $K^{(n)}$ while the $M^{(n)}_i$ transform as a vector under $SO(d)$. The generators $L^{(n)}$ and $M^{(n)}_i$ span the infinite dimensional Galilean conformal algebra observed in \cite{Bagchi:2009my} (see also \cite{Martelli:2009uc}). It was shown in \cite{Bagchi:2014ysa} that this is a symmetry of the electric and magnetic theory for $d=3$. Here we see that this algebra exists for any dimension. Further the actual symmetry algebra of the equations of motion of the electric and magnetic limit is larger than the one of \cite{Bagchi:2014ysa} because it includes the $U(1)$ current algebra spanned by the $K^{(n)}$ generators. The action of the $L^{(n)}$,  $M^{(n)}_i$ and $K^{(n)}$ on the fields appearing in the electric and magnetic limits can be inferred from \eqref{eq:ElimitSym} and \eqref{eq:MlimitSym}. In both cases there is also an additional symmetry corresponding to an overall rescaling of all the fields whose parameters are $\gamma$ and $\bar\gamma$. The subalgebra of $L^{(n)}$ and $K^{(n)}$ is the same infinite dimensional algebra observed in the context of warped CFTs \cite{Hofman:2011zj,Detournay:2012pc}. Here we did not study any possible central charges. 

\subsubsection*{On- and off-shell GED} 

We will now add the equation of motion \eqref{eq:extraeom} and demand it is invariant under \eqref{eq:ElimitSym}. This leads to severe constraints on the scalars $\xi^t$ and $\Omega$. The transformations leaving the equations of motion of the action \eqref{eq:flatGED} invariant are
\begin{eqnarray}
\delta\varphi & = & \xi^t\partial_t\varphi+\xi^k\partial_k\varphi+\left[(d-2)ct+(d-2)\mu+\gamma\right]\varphi\,,\\
\delta\tilde\varphi & = & \xi^t\partial_t\tilde\varphi+\xi^k\partial_k\tilde\varphi+\alpha_3(t)\varphi-a_k\partial_t\xi^k+\left[-(d-4)ct+2\lambda+(d-4)\mu+\gamma\right]\tilde\varphi\,,\\
\delta a_i & = & \xi^t\partial_t a_i+\xi^k\partial_k a_i+a_k\partial_i\xi^k-\varphi\partial_t\xi^i+\left[\lambda+(d-4)\mu+\gamma\right] a_i\,,\label{eq:onshellSym}
\end{eqnarray}
where $\xi^t$ and $\xi^i$ are given by
\begin{eqnarray}
\xi^t & = & \zeta^t+\lambda t-\frac{1}{2}(d-4)ct^2\,,\\
\xi^i & = & \zeta^i(t)+\lambda^i{}_j x^j+\mu x^i+ctx^i=\zeta^i(t)+\lambda^i{}_j x^j+\partial_t\xi^t x^i+\bar\Omega x^i\,,
\end{eqnarray}
and where $\bar\Omega$ is defined in \eqref{eq:defOmega} and given by $\bar\Omega=\mu-\lambda+(d-3)ct$. With the exception of $\alpha_3(t)$ and $\zeta^i(t)$ all parameters are constants. The parameters $\lambda$ and $\mu$ are two independent scaling parameters corresponding to the fact that we can scale time and space independently accompanied by appropriate rescalings of the fields. The parameter $\gamma$ corresponds to a rescaling of all the fields that follows from the fact that the equations of motion are linear in the fields. The algebra is infinite dimensional  because of the time dependent translations $\zeta^i(t)$. The generators of this infinite dimensional algebra are
\begin{eqnarray}
&&M^{(n)}_i=t^{n+1}\partial_i\,,\qquad H=\partial_t\,,\qquad D_1=t\partial_t\,,\qquad D_2=x^i\partial_i\,,\qquad J_{ij}=x^i\partial_j-x^j\partial_i\,,\nonumber\\
&&K=-\frac{1}{2}(d-4)t^2\partial_t+tx^i\partial_i\,.
\end{eqnarray}
The nonzero commutators are given by
\begin{eqnarray}
&&[K\,,M^{(n)}_i]=-\frac{1}{2}\left(d-2+(d-4)n\right)M^{(n+1)}_i\,,\qquad [H\,,M^{(n)}_i]=(n+1)M^{(n-1)}_i\,,\nonumber\\
&&[D_1\,,M^{(n)}_i]=(n+1)M^{(n)}_i\,,\qquad [K\,,H]=(d-4)D_1-D_2\,,\qquad [D_1\,,K]=K\,,\nonumber\\
&& [D_1\,,H]=-H\,.
\end{eqnarray}
The parameter $c$ corresponds for $d=3$ to a special conformal transformation. In fact transformations for which $\mu=\lambda$ and $d=3$ so that $\bar\Omega=0$ contain the finite dimensional Galilean conformal algebra consisting of the generators $H$, $D_1+D_2$, $K$, $J_{ij}$, $M_i^{(-1)}$, $M_i^{(0)}$ and $M_i^{(1)}$.

Finally we will determine which of these on-shell symmetries leave the GED action invariant. Invariance of \eqref{eq:flatGED} is obtained if the Lagrangian density obeys $\delta\mathcal{L}=\partial_\mu\left(\xi^\mu\mathcal{L}\right)$. This leads to the following restrictions
\begin{equation}
\alpha_3=0\,,\qquad \zeta^i(t)=\zeta^i+v^i t\,,\qquad c=0\,,\qquad\gamma=-\frac{1}{2}\lambda-\frac{1}{2}(d-4)\mu\,.
\end{equation}
Hence the off-shell symmetries of GED are
\begin{eqnarray}
\delta\varphi & = & \xi^t\partial_t\varphi+\xi^k\partial_k\varphi+\left[-\frac{1}{2}\lambda+\frac{d}{2}\mu\right]\varphi\,,\\
\delta\tilde\varphi & = & \xi^t\partial_t\tilde\varphi+\xi^k\partial_k\tilde\varphi-a_k\partial_t\xi^k+\left[\frac{3}{2}\lambda+\frac{1}{2}(d-4)\mu\right]\tilde\varphi\,,\\
\delta a_i & = & \xi^t\partial_t a_i+\xi^k\partial_k a_i+a_k\partial_i\xi^k-\varphi\partial_t\xi^i+\left[\frac{1}{2}\lambda+\frac{1}{2}(d-4)\mu\right] a_i\,,\label{eq:offshellSym}
\end{eqnarray}
where $\xi^t$ and $\xi^i$ are given by
\begin{eqnarray}
\xi^t & = & \zeta^t+\lambda t\,,\\
\xi^i & = & \zeta^i+v^it+\lambda^i{}_j x^j+\mu x^i\,.
\end{eqnarray}
The translational and rotational symmetries are obvious. The Galilean invariance has already been discussed in section \ref{sec:NRlimit}. The finite version of the scale symmetries are
\begin{eqnarray}
&&t\rightarrow\lambda t\,,\qquad\varphi\rightarrow\lambda^{1/2}\varphi\,,\qquad\tilde\varphi\rightarrow\lambda^{-3/2}\tilde\varphi\,,\qquad a_i\rightarrow\lambda^{-1/2}a_i\,,\nonumber\\
&&x^i\rightarrow\mu x^i\,,\qquad\varphi\rightarrow\mu^{-d/2}\varphi\,,\qquad\tilde\varphi\rightarrow\mu^{-(d-4)/2}\tilde\varphi\,,\qquad a_i\rightarrow\mu^{-(d-2)/2}a_i\,.
\end{eqnarray}
Note that the scaling weight of $a_i$ gets a contribution from the $a_k\partial_i\xi^k$ term in \eqref{eq:onshellSym}. These symmetries form a Lie algebra consisting of the Galilei algebra and two dilatations generators. 

The generators of the symmetries of the action are
\begin{eqnarray}
&&H=\partial_t\,,\qquad P_i=\partial_i\,,\qquad J_{ij}=x^i\partial_j-x^j\partial_i\,,\qquad G_i=t\partial_i\,,\nonumber\\
&&D_1=t\partial_t\,,\qquad D_2=x^i\partial_i\,.
\end{eqnarray}
The first line gives the Galilean algebra and the second line are the two dilatation generators. The nonzero commutators of $D_1$ and $D_2$ with the Galilean algebra are
\begin{eqnarray}
&&\left[D_1\,, H\right]=-H\,,\qquad \left[D_1\,,G_i\right]=G_i\,,\nonumber\\
&&\left[D_2\,, P_i\right]=-P_i\,,\qquad \left[D_2\,,G_i\right]=-G_i\,,
\end{eqnarray}
while $D_1$ and $D_2$ commute with each other. The off-shell GED theory is an example of a Galilean field theory that has a Lifshitz scale invariance for any value of $z$. This is a very attractive property because it means that we can couple GED to all matter theories with any critical exponent $z$ without breaking the scale symmetry of the matter theory. 

We thus see that the off-shell theory has fewer symmetries than the on-shell theory. In particular GED ceases to be conformal for $d=3$ off-shell. This is different from Maxwell's theory in $3+1$ dimensions. This can be understood from the fact the GED Lagrangian is the contraction of Maxwell coupled to a free scalar with a shift symmetry (see Section \ref{sec:NRlimit}). It is well-known that free scalars with a shift symmetry are off-shell scale invariant theories that are not conformally invariant. This is  because the total derivative term that one would have to add to the Lagrangian to improve the energy-momentum tensor to one that is traceless breaks the shift symmetry.

\section{Coupling to TNC geometry}\label{sec:TNC}

We will study the coupling of the three different limit theories discussed in section \ref{sec:NRlimit} to an arbitrary curved background described by torsional Newton--Cartan geometry (TNC).  We will start with the coupling of the GED limit to TNC curved space. This case is simpler because it admits a Lagrangian description. We will then consider the electric and magnetic limits and conclude that in order to have local equations of motion the space-time geometry needs to be restricted. In particular the geometry will be twistless torsional Newton--Cartan (TTNC) whose definition we will review. 

\subsubsection*{Summary of TNC geometry}\label{app:TNC}

Here we briefly review TNC geometry and our conventions following~\cite{Hartong:2014oma,Hartong:2014pma,Bergshoeff:2014uea,Hartong:2015wxa,Hartong:2015zia} (see also \cite{Bekaert:2014bwa} for further details of TNC connections). 

A Torsional Newton--Cartan background in $d+1$ dimensions is given by a set of one forms (vielbeins) $\left(\tau_\mu\,, e_\mu^a\right)$ where $a=1\ldots d$ and a one form $M_\mu$.  The inverse vielbeins $v^\mu$ and $e^\mu_a$ are defined through
\begin{equation}
v^\mu e_\mu^a = 0\,,\qquad v^\mu\tau_\mu = -1\,,\qquad e^{\mu}_a\tau_\mu = 0\,,\qquad e^{\mu}_ae_{\mu}^b = \delta^b_a\,.
\end{equation}
The determinant of the square matrix $\left(\tau_\mu\,, e_\mu^a\right)$ is denoted by $e$. The vielbeins can be used to construct a degenerate ``spatial metric'' $h_{\mu\nu}=\delta_{ab}e^a_\mu e^b_\nu$ and similarly $h^{\mu\nu}=\delta^{ab}e_a^\mu e_b^\nu$. When $\tau_\mu$ is surface orthogonal the geometry is referred to as Twistless TNC and $h_{\mu\nu}$ is a Riemannian metric on the surfaces orthogonal to $\tau_\mu$.

Besides transforming under diffeomorphisms as usual, the one forms  $\tau_\mu$, $e_\mu^a$ and $M_\mu$ transform under various local transformations: Galilean boosts with $\lambda_a$ as local parameter, local $SO(d)$ rotations parametrized by $\lambda_{a b}=-\lambda_{b a}$ and $U(1)_\sigma$ gauge transformations parametrized~by~$\sigma$,
\begin{eqnarray}
&&\delta \tau_\mu = 0\,, \qquad\quad\delta e_\mu^a = \tau_\mu\lambda^a + \lambda^{a}{}_b e_{\mu}^{b}\cr 
&& \delta v^\mu = \lambda^a e^\mu_a\,, \quad\;\delta e^\mu_a =  \lambda_{a}{}^b e^{\mu}_{b} \,,\cr
&&\delta M_\mu = \lambda_a e_{\mu}^a +\partial_\mu\sigma\,.
\end{eqnarray}

The inverse vielbein $e^\mu_a$ and $h^{\mu\nu}$ are invariant under local Galilean transformations. It is useful to define other objects with this property $\hat v^\mu$,~$\hat e_\mu^a$,~$\hat h_{\mu\nu}$ and $\tilde\Phi$\footnote{ $\tilde\Phi$ is related to the Newtonian potential \cite{Bergshoeff:2014uea,Hartong:2015wxa}~.}  via
\begin{eqnarray}
&&\hat v^\mu = v^\mu-e^{\mu}_a M_\nu e^{\nu a}\,,\qquad \hat e_\mu^a = e_\mu^a-M_\nu e^{\nu a}\tau_\mu\,,\cr
&&\hat h_{\mu\nu}=h_{\mu\nu}-M_\mu\tau_\nu-M_\nu\tau_\mu\,,\quad \tilde\Phi=-v^\mu M_\mu+\frac{1}{2}h^{\mu\nu}M_\mu M_\nu\,.
\end{eqnarray}
These objects satisfy the relations:
\begin{eqnarray}
\hat v^\mu\hat e_\mu^a = 0\,,\qquad \hat v^\mu\tau_\mu = -1\,,\qquad e^{\mu}_a\tau_\mu = 0\,,\qquad e^{\mu}_a\hat e_{\mu}^b = \delta^b_a\,.
\end{eqnarray}

\subsection{GED on a TNC background}

We introduce the $U(1)$ gauge field $\bar A_\mu$ and the scalar field $\varphi$ which transform as follows under local Galilean boosts 
\begin{equation}
\label{eq:bargauge}
\delta \bar A_\mu=\varphi\, e_{\mu}^a \lambda_a \,,\quad \delta  \varphi=0\,,
\end{equation}
Under local $U(1)_\sigma$ transformations and $SO(d)$ rotations $\bar A_\mu$ and $\varphi$ are both invariant. The gauge field $\bar A_\mu$ has the usual gauge redundancy: $\bar A_\mu\sim \bar  A_\mu+\partial_\mu \Lambda$.

We can write $\bar A_\mu=a_\mu-\tilde\varphi\tau_\mu$ where $v^\mu a_\mu=0$. We find that $a_\mu$ and $\varphi$ transform as follows under local Galilean boosts and gauge transformations:
\begin{equation}
\label{eq:barcompt}
a_\mu\sim a_\mu + \tau_\mu v^\nu \partial_\nu \Lambda\,,\quad \delta  a_\mu=\varphi\, e_{\mu}^a \lambda_a +\tau_\mu a_\nu e^\nu_a \lambda^a\,,\quad \tilde \varphi\sim \tilde \varphi+v^\nu \partial_\nu\Lambda \,,\quad  \delta \tilde\varphi=a_\nu e^\nu_a \lambda^a\,.
\end{equation}
In the flat limit of the TNC geometry we have $\tau_\mu=\delta_\mu^t$, $~e^\mu_a=\delta^\mu_a$, $~v^\mu=-\delta^\mu_t$ and $e^a_\mu=\delta^a_\mu$. The flat space GED fields are given by $a_i=a_\mu e^\mu_i\,,~\tilde\varphi$, and $\varphi$. Indeed these fields transform as in \eqref{eq:GEDboosts} under infinitesimal Galilean boosts parametrized by constant $\lambda_a$. 

We will define  the following field strength for $\bar A_\mu$ 
$$\bar F_{\mu\nu}=\partial_\mu\bar A_\nu-\partial_\nu\bar A_\mu-\varphi\left(\partial_\mu M_\nu-\partial_\nu M_\mu\right)~.$$ 
We can then write down an action for GED coupled to an arbitrary TNC background as follows
\begin{equation}\label{eq:GED2}
S_{\text{GED}}=\int d^{d+1}x \,e\, \left(-\frac{1}{4}h^{\mu\rho}h^{\nu\sigma}\bar F_{\mu\nu}\bar F_{\rho\sigma}-h^{\mu\nu}v^\rho\bar F_{\rho\nu}\partial_\mu\varphi+\frac{1}{2}\left(v^\mu\partial_\mu\varphi\right)^2\right)\,.
\end{equation}
In this form the action is manifestly invariant under diffeomorphisms and $U(1)_\sigma$ transformations. It is also invariant under local Galilean boosts and rotations.

Alternatively we can forgo manifest $U(1)_\sigma$ invariance and rewrite the action in terms of Galilean invariant objects. Indeed we can define a new gauge potential 
$$A_\mu=\bar A_\mu-\varphi M_\mu=a_\mu-\tau_\mu \tilde \varphi -\varphi M_\mu~,$$
 which is inert under local Galilean boosts and transforms under $U(1)_\sigma$ as $\delta A_\mu=-\varphi\partial_\mu\sigma$. In terms of $A_\mu$ the action \eqref{eq:GED2} is given by
\begin{equation}\label{eq:GED}
S_{\text{GED}}\!=\!\int d^{d+1}x \,e\left(\!-\frac{1}{4}h^{\mu\rho}h^{\nu\sigma}F_{\mu\nu}F_{\rho\sigma}-h^{\mu\nu}\hat v^\rho F_{\rho\nu}\partial_\mu\varphi-\tilde\Phi h^{\mu\nu}\partial_\mu\varphi\partial_\nu\varphi+\frac{1}{2}\left(\hat v^\mu\partial_\mu\varphi\right)^2\right)\,,
\end{equation}
where $F_{\mu\nu}=\partial_\mu A_\nu-\partial_\nu A_\mu$. 

By varying the GED action \eqref{eq:GED} we obtain the equations of motion
\begin{equation}
\partial_\mu\left(e\,\tilde F^{\mu\nu}\right)=0\,,\qquad\partial_\mu\left(e\, \tilde G^\mu\right)=0\,
\end{equation}
where $\tilde F^{\mu\nu}$ and $\tilde G^\mu$ are defined as
\begin{eqnarray}
\label{eq:eomgedtnc}
\tilde F^{\mu\nu} & = & h^{\mu\rho}h^{\nu\sigma}F_{\rho\sigma}+\left(\hat v^\mu h^{\nu\rho}-\hat v^\nu h^{\mu\rho}\right)\partial_\rho\varphi\,,\\
\tilde G^\mu & = & h^{\mu\nu}\hat v^\rho F_{\rho\nu}+2\tilde\Phi h^{\mu\nu}\partial_\nu\varphi-\hat v^\mu\hat v^\nu\partial_\nu\varphi\,.
\end{eqnarray}
Note that $\tilde F^{\mu\nu}$ is invariant under both $U(1)_\Lambda$ and $U(1)_\sigma$ transformations while $\tilde G^\mu$ is $U(1)_\Lambda$ invariant but transforms under $U(1)_\sigma$ as $\delta_\sigma \tilde G^\mu=\tilde F^{\mu\nu}\partial_\nu\sigma$. Hence the equation of motion $\partial_\mu\left(e\,\tilde G^\mu\right)=0$ is $U(1)_\sigma$ invariant by virtue of the other equation of motion $\partial_\mu\left(e\tilde F^{\mu\nu}\right)=0$.

We remark that the linearized version of the GED action coupled to TNC was also obtained in \cite{Festuccia:2016awg} via
the Noether procedure. This paper also shows that in theories with massless Galilean symmetries, of which GED is an example,
the TNC vector $M_\mu$ couples to a topological current. We refer the reader to this paper for 
the explicit form of this topological current for GED, along with the other (improved) currents.

\subsubsection*{Null reduction of Maxwellian electromagnetism}\label{subsec:nullredcr}

The GED action on TNC geometry can also be obtained by null reduction of Maxwellian electromagnetism in one dimension higher. Consider the Maxwell action coupled to a background Lorentzian metric $\gamma_{AB}$,
\begin{equation}\label{eq:Mcur}
S=-\int d^{d+2}x\sqrt{-\gamma}\frac{1}{4}F_{AB}F^{AB}\,,
\end{equation}
where $F=dA$. 
We can now restrict the background metric to possess a null isometry, which in suitably chosen coordinates is generated by $\partial_u$
\begin{eqnarray}
\label{eq:invmet}
ds^2 & = & \gamma_{AB}dx^A dx^B = 2\tau_\mu dx^\mu\left(du-M_\nu dx^\nu\right)+h_{\mu\nu}dx^\mu dx^\nu\,,\\
\label{eq:invmeta}
\sqrt{-\gamma} & = & e\,,\qquad\gamma^{uu} = 2\tilde\Phi\,,\qquad\gamma^{u\mu}=-\hat v^\mu\,,\qquad \gamma^{\mu\nu}=h^{\mu\nu}\,,\qquad A_u = \varphi\,.
\end{eqnarray}
This form of the metric is preserved by the following changes of coordinates: 
\begin{itemize}
\item{}{$x'^\mu=x'^\mu(x)$  identified with diffeomorphisms in the lower dimensional TNC geometry.}
\item{}{$u'=u+\sigma(x)$ that give rise to $U(1)_\sigma$ transformations. }
\end{itemize}
The Galilean invariant objects of section \ref{app:TNC} correspond to the components of $\gamma_{AB}$ as in \eqref{eq:invmeta}. 

We want to reduce \eqref{eq:Mcur} along the null isometry. For this we will restrict the gauge field $A_M$ and the $U(1)$ gauge parameter $\Lambda$ to be invariant along $\partial_u$. We can then use $A_M=(A_u, A_\mu)$ to define two lower dimensional fields. The first one $\varphi\equiv A_u$ is a gauge invariant scalar. The second one $\bar A_\mu\equiv A_\mu+\varphi M_\mu$ is a lower dimensional gauge field that is invariant under $U(1)_\sigma$ transformations. This procedure leads directly to the action \eqref{eq:GED} for GED coupled to a TNC background. 

\subsection{The magnetic and electric theory on a TTNC background}\label{subsec:magTNC}

We can obtain the equations of motion for the magnetic theory by solving for $\varphi$ in the GED equations of motion  \eqref{eq:eomgedtnc}. In parallel to flat space we can consider
\begin{equation}
\tau_\nu \partial_\mu\left(e\,\tilde F^{\mu\nu}\right)=0\quad \Rightarrow\quad  {1\over e}\partial_\mu(e h^{\mu\nu} \partial_\nu \varphi)-\hat v^\mu\left(\partial_\mu\tau_\nu-\partial_\nu\tau_\mu\right)h^{\nu\rho}\partial_\rho\varphi={1\over 2} (\partial_\mu \tau_\nu-\partial_\nu \tau_\mu)h^{\mu\rho}h^{\nu\sigma}F_{\rho\sigma}~.
\end{equation}
In general solving this equation for $\varphi$ and substituting back into the remaining equations of motion would not result in local expressions. However if $\tau\wedge d\tau=0$ the right hand side of the equation above vanishes and $\varphi=0$ is a solution.
The equations of motion for the magnetic theory on a TTNC background can then be written as:
\begin{eqnarray}
\label{eq:ttncmag}
&&\partial_\mu\left(e h^{\mu\rho}h^{\nu\sigma}(\partial_{\rho}\bar A_{\sigma}-\partial_{\sigma}\bar A_{\rho})\right)  =  0\,,\\
&&\partial_\mu\left(e  h^{\mu\sigma}\hat v^\rho(\partial_{\rho}\bar A_{\sigma}-\partial_{\sigma}\bar A_{\rho})\right) = 0\,.
\end{eqnarray}
Because of \eqref{eq:bargauge} and the fact that $\varphi =0$ it follows that $\bar A_\mu$ is now inert under local  Galilean boosts.

Turning to the electric theory, we can use the second GED equation \eqref{eq:eomgedtnc} to solve for $\tilde \varphi$. In parallel with what happens in flat space the first equation in \eqref{eq:eomgedtnc} would then describe the electric theory coupled to curved space. This in general will result in nonlocal equations for the electric fields. However note that defining $A^{\text{el}}_\mu=A_\mu+\tau_\mu \tilde \varphi =a_\mu-\varphi M_\mu$ we can write
\begin{equation}
h^{\mu\rho}h^{\nu\sigma}F_{\rho\sigma}=h^{\mu\rho}h^{\nu\sigma}(\partial_\rho A^{\text{el}}_\sigma -\partial_\sigma A^{\text{el}}_\rho)-\tilde\varphi h^{\mu\rho}h^{\nu\sigma}\left(\partial_\rho\tau_\sigma-\partial_\sigma\tau_\rho\right)\,.
\end{equation}
When the geometry is twistless the term proportional to $\tilde \varphi$ in the above equation vanishes. As a consequence on a twistless background $\tilde \varphi$ does not appear in the GED equation of motion $\partial_\mu\left(e\tilde F^{\mu\nu}\right) =  0$. We conclude that on a TTNC background the equations of motion for the electric fields $a_\mu$ and $\varphi$ are still local after solving for $\tilde \varphi$ and are given by:
\begin{eqnarray}
\partial_\mu\left(e F_{\text{el}}^{\mu\nu}\right) & = & 0\,,\qquad F_{\text{el}}^{\mu\nu}= h^{\mu\rho}h^{\nu\sigma}(\partial_\rho A^{\text{el}}_\sigma -\partial_\sigma A^{\text{el}}_\rho)+\left(\hat v^\mu h^{\nu\rho}-\hat v^\nu h^{\mu\rho}\right)\partial_\rho\varphi\,,
\end{eqnarray}
For instance contracting $\partial_\mu\left(e\tilde F_{\text{el}}^{\mu\nu}\right)=0$ with $\tau_\nu$ we find
\begin{equation}
e^{-1}\partial_\mu\left(eh^{\mu\nu}\partial_\nu\varphi\right)-\hat v^\mu h^{\nu\rho}\left(\partial_\mu\tau_\nu-\partial_\nu\tau_\mu\right)\partial_\rho\varphi=0\,,
\end{equation}
which is the TTNC generalization of the first equation in \eqref{eq:Elimit}.

The field $\varphi$ is inert under local Galilean boosts, gauge transformations and $U(1)_\sigma$ transformations. The field $a_\mu$, which satisfies $v^\mu a_\mu=0$ is also invariant under $U(1)_\sigma$ but  transforms under Galilean boosts and gauge transformations as follows:
$$\delta a_\mu=\varphi e_\mu^a\lambda_a+\lambda^a e^\nu_a a_\nu \tau_\mu~,\qquad \delta_\Lambda a_\mu=\partial_\mu\Lambda+\tau_\mu v^\nu \partial_\nu \Lambda~.$$

\subsection{Coupling to charged matter}\label{sec:matter}

The GED action \eqref{eq:GED2} or \eqref{eq:GED} has a local $U(1)_\Lambda\times U(1)_\sigma$ symmetry. We will first construct a minimal coupling to charged scalar fields that respects this symmetry and then we will generalize it by inclusion of non-minimal couplings. We will obtain the minimal coupling as the null reduction of scalar QED in one dimension higher.

The $(d+2)$-dimensional theory is
\begin{equation}
S=\int d^{d+2}x\sqrt{-\gamma}\left(-\gamma^{AB}D_A\psi D_B\psi^\star-\frac{1}{4}F_{AB}F^{AB}\right)\,,
\end{equation}
where $D_A\psi=\partial_A\psi-iqA_A\psi$ and $q$ is the electric charge. The metric has the same form as in \eqref{eq:invmet}. Writing $\psi=e^{imu}\phi$ with $\phi$ independent of $u$  and reducing along $u$ we obtain:
\begin{eqnarray}
S & = & \int d^{d+1}x e\left(-i(m-q\varphi)\phi^\star\hat v^\mu D_\mu\phi+i(m-q\varphi)\phi\hat v^\mu D_\mu\phi^\star-h^{\mu\nu}D_\mu\phi D_\nu\phi^\star\right.\nonumber\\
&&\left.-2(m-q\varphi)^2\tilde\Phi\phi\phi^\star\right)+S_{\text{GED}}\,.\label{eq:dimredminmodel}
\end{eqnarray}
Here $D_\mu= \partial_\mu - i e A_\mu~.$ The scalar field $\phi$ is inert under local Galilean boosts but transforms under $U(1)_\Lambda$ and $U(1)_\sigma$ as $\delta \phi= i(q \Lambda -m \sigma)\phi$. The invariance of the action under boosts is explicit; in order to make the $U(1)_\sigma$ invariance manifest we can rewrite the action as
\begin{equation}
\!S\!=\! \int d^{d+1}x e\left(-i\left(m-q\varphi\right)\phi^\star v^\mu\hat D_\mu\phi+i\left(m-q\varphi\right)\phi v^\mu\hat D_\mu\phi^\star-h^{\mu\nu}\hat D_\mu\phi\hat D_\nu\phi^\star \right)\!+\!S_{\text{GED}}\,,
\end{equation}
where 
\begin{equation}
\hat D_\mu\phi=D_\mu\phi+i\left(m-q\varphi\right)M_\mu\phi=\partial_\mu\phi-iqa_\mu\phi+iq\tilde\varphi\tau_\mu\phi+imM_\mu\phi\,.
\end{equation}
The equations of motion for the gauge fields are given by:
\begin{eqnarray}
\label{eq:eomX}
&&e^{-1} \partial_\mu\left(e\,\tilde F^{\mu\nu}\right)=J^\nu\,,\qquad e^{-1}\partial_\mu\left(e\, \tilde G^\mu\right)=\tilde \rho\,\\
&&J^\mu  = 2q(m-q\varphi)\phi\phi^\star\hat v^\mu-iqh^{\mu\nu}\left(\phi D_\nu\phi^\star-\phi^\star D_\nu\phi\right)\,,\label{eq:eomXa}\\
&&\tilde\rho  =  iq\hat v^\mu\left(\phi D_\mu\phi^\star-\phi^\star D_\mu\phi\right)-4q(m-q\varphi)\tilde\Phi\phi\phi^\star\,.
\end{eqnarray}
They can be shown to be invariant under $U(1)_\sigma$ using  $\delta_\sigma \tilde \rho=J^\mu\partial_\mu \sigma$.
The equation of motion for  the scalar field $\phi$ reads
\begin{equation}
-2i(m-q\varphi)\hat v^\mu D_\mu\phi-i\phi e^{-1}\partial_\mu\left(e(m-q\varphi)\hat v^\mu\right)+e^{-1}D_\nu\left(eh^{\mu\nu}D_\mu\phi\right)-2(m-q\varphi)^2\tilde\Phi\phi=0\,;
\end{equation}
it can be used to check that $\partial_\mu(e J^\mu)=0$ as required by the first equation in \eqref{eq:eomX}.

Next we will consider the electric theory coupled to matter in curved space. In the electric limit the scalar $\phi$ does not transform under $U(1)_\Lambda$. The equations of motion in flat space \eqref{eq:Elimsqed}  are  extended to a TTNC spacetime as \begin{eqnarray}
\label{eq:eomelX}
&&e^{-1} \partial_\mu\left(e\, F_{\text{el}}^{\mu\nu}\right)=J^\nu\,,\qquad J^\mu  = 2q m \phi\phi^\star\hat v^\mu-iqh^{\mu\nu}\left(\phi \partial_\nu\phi^\star-\phi^\star \partial_\nu\phi\right)\,,\\
&&2i m(\hat v^\mu \partial_\mu-i q \varphi)\phi+i m\phi e^{-1}\partial_\mu\left(e\hat v^\mu\right)-e^{-1}\partial_\nu\left(e h^{\mu\nu}\partial_\mu\phi\right)+2m^2\tilde\Phi\phi=0\,. \nonumber
\end{eqnarray}

The magnetic theory in flat space can be coupled to sources as in \eqref{eq:generalsourcedmagnetic}. In a curved TTNC background the sources modify \eqref{eq:ttncmag} as follows: 
\begin{eqnarray}
\label{eq:eommagX}
\!\!\!\!e^{-1} \partial_\mu\left(e h^{\mu\rho}h^{\nu\sigma}(\partial_{\rho}\bar A_{\sigma}-\partial_{\sigma}\bar A_{\rho})\right)=J^\nu\,,\quad e^{-1} \partial_\mu \left( e  h^{\mu\sigma}\hat v^\rho(\partial_{\rho}\bar A_{\sigma}-\partial_{\sigma}\bar A_{\rho})\right)=\rho- M_\nu J^\nu \,.
\end{eqnarray}
Here the current $J^\mu$ satisfies $\tau_\mu J^\mu=0$ and is conserved $\partial_\mu (e J^\mu)=0$. Under $U(1)_\sigma$ both $\rho$ and $J^\mu$ are invariant while under local Galilean boosts we have $\delta J^\mu=0$ and $\delta \rho= J^\nu e_\nu^a \lambda_a$.

As in flat space a charged scalar field gives rise to source terms for the magnetic theory that are not of the most general form: $J^\mu=0$ and $\rho=-q \phi \phi^\star$. The equation of motion of the charged scalar reads
\begin{eqnarray}
\label{eq:scalmagX}
&&-2im\hat v^\mu D_\mu\phi-i m \phi e^{-1}\partial_\mu\left(e \hat v^\mu\right)+e^{-1}D_\nu\left(eh^{\mu\nu}D_\mu\phi\right)-2 m^2\tilde\Phi\phi=0\,,
\end{eqnarray}
where $D_\mu \phi= \partial_\mu \phi -i q \bar A_\mu \phi= \partial_\mu \phi -i q a_\mu \phi+i q \tau_\mu \tilde \varphi \phi$.
This equation of motion can be obtained from the action 
\begin{equation}\label{eq:magneticSch2}
S=\int d^{d+1}x e\left(-im\phi^\star v^\mu\mathcal{D}_\mu\phi+im\phi v^\mu\mathcal{D}_\mu\phi^\star-h^{\mu\nu}\mathcal{D}_\mu\phi\mathcal{D}_\nu\phi^\star\right)\,,
\end{equation}
where $\mathcal{D}_\mu\phi=\partial_\mu\phi-iq\bar A_\mu\phi+imM_\mu\phi$.  If the gauge potential of the magnetic theory $\bar A_\mu$ is a fixed background field then this action is Galilean invariant in a general TNC geometry and the restriction to TTNC which was required for \eqref{eq:eommagX} to make sense is no longer needed.
In this case the action \eqref{eq:magneticSch2} agrees with the one presented in \cite{Jensen:2014aia} where $\bar A_\mu$ is absorbed into $M_\mu$. However in general this is not possible. If we consider several copies of $\phi$, say $\phi_1$ and $\phi_2$ with charges and masses $(q_1,m_1)$ and $(q_2,m_2)$ respectively and such that $q_1/m_1\neq q_2/m_2$ the couplings to $\bar A_\mu$ and $M_\mu$ are no longer proportional and we cannot absorb $\bar A_\mu$ into $M_\mu$. 

For completeness we also consider the Lagrangian for a charged point particle coupled to GED \eqref{eq:sged} which can  be extended to a curved TNC background:  
\begin{equation}
S=\int d\lambda\left(\frac{1}{2}(m-q\varphi)\frac{\bar h_{\mu\nu}\dot X^\mu\dot X^\nu}{\tau_\rho\dot X^\rho}+qA_\mu\dot X^\mu\right)\,,
\end{equation}
where $\hat h_{\mu\nu}$ is defined in section \ref{app:TNC} and dots denote derivatives with respect to $\lambda$. The Galilean boost invariance is manifest. To see the invariance under $U(1)_\sigma$ it is useful to go from $\bar h_{\mu\nu}$ to $h_{\mu\nu}$. This action is the  generalization to the charged case of the action given in \cite{Kuchar:1980tw,Bergshoeff:2014gja,Hartong:2015wxa}. 

\subsubsection*{Non minimal couplings}

An interesting example of a non-minimal model is obtained by the null reduction of the following (Pauli coupling) relativistic action
\begin{equation}
S=\int d^{d+2}x\sqrt{-\gamma}\left(-\gamma^{AB}D_A\psi D_B\psi^\star-\frac{1}{4}\left(1+g\psi\psi^\star\right)F_{AB}F^{AB}\right)\,,
\end{equation}
where $g$ is a coupling constant. After null reduction we obtain
\begin{eqnarray}
S & = & \int d^{d+1}x e\left[-i(m-q\varphi)\phi^\star\hat v^\mu D_\mu\phi+i(m-q\varphi)\phi\hat v^\mu D_\mu\phi^\star-h^{\mu\nu}D_\mu\phi D_\nu\phi^\star-2(m-q\varphi)^2\tilde\Phi\phi\phi^\star\right.\nonumber\\
&&\left.+\left(1+g\phi\phi^\star\right)\left(-\frac{1}{4}h^{\mu\rho}h^{\nu\sigma}F_{\mu\nu}F_{\rho\sigma}-h^{\mu\nu}\hat v^\rho F_{\rho\nu}\partial_\mu\varphi-\tilde\Phi h^{\mu\nu}\partial_\mu\varphi\partial_\nu\varphi+\frac{1}{2}\left(\hat v^\mu\partial_\mu\varphi\right)^2\right)\right]\,.
\end{eqnarray}
We can also generalize the higher dimensional model by adding a potential term $-\sqrt{-\gamma}V(\psi\psi^\star)$ which  reduces to $-eV(\phi\phi^\star)$. 

\section{Outlook} \label{sec:outlook}

We conclude by mentioning a number of interesting directions for further work. 

First of all, we recall the directions mentioned in the introduction as  motivations for the present work which will be worthwhile to study given
our results. These include examining the appearance of non-relativistic electrodynamic fields as background sources in Lifshitz holography with
extra bulk Maxwell fields, e.g. by adding a Maxwell field to the EPD model of \cite{Christensen:2013lma,Christensen:2013rfa,Hartong:2014oma}. 
Another application is to consider dynamical (T)TNC gravity \cite{Hartong:2015zia,Hartong:2016yrf} coupled to 
GED as a holographic bulk gravity theory.  Furthermore, it would be interesting to see whether one can construct
a supersymmetric version of GED. 

We also  note that the scalar field $\varphi$ (mass potential) in the GED action is invariant under all the relevant symmetries, including
Galilean boosts, the Bargmann $U(1)_\sigma$ and gauge $U(1)_\Lambda$. Consequently we can add potential terms such
as $V(\varphi)$ since they preserve the symmetries\footnote{We can also add a term like $-\tfrac{x}{2}(\partial_i\varphi)^2$ but this can be removed by redefining $\tilde\varphi$ in \eqref{eq:flatGED} to $\tilde\varphi-\tfrac{x}{2}\varphi$.}. It would be interesting to study the effect of these terms on the symmetries
and couplings that we have found.  
Another generalization would involve adding higher spatial derivative terms, as seen in Lifshitz scalar field theories. 
Furthermore, one could examine Hodge duality and electromagnetic duality for non-relativistic electrodynamics,
including the coupling of magnetic monopoles to GED. 

We have seen in this paper that performing null reductions on relativistic theories is a powerful tool to obtain consistent non-relativistic
theories and provides a simple way to derive the couplings to non-relativistic backgrounds.
It would thus be interesting to apply this to relativistic fields of spin $s$ (massive or massless) and Yang-Mills theories,
and compare to the works \cite{Fuini:2015yva,Bagchi:2015qcw,Bergshoeff:2015sic} in which various non-relativistic cases are considered. For example (twisted) null reduction of $\mathcal{N}=4$ SYM plays an important role in the description of the boundary theory of 4-dimensional $z=2$ Lifshitz space-times following \cite{Christensen:2013rfa}.
In another direction, it would be interesting to add a Chern--Simons coupling to Einstein-Maxwell in five dimensions and determine 
the resulting terms in four-dimensional GED after null reduction. 

Another extension of the present work is to consider the Proca theory. Using null reduction of the $D+1$ dimensional
Proca term $\mu^2 A_BA^B/2$ 
it is not difficult to see that this adds the terms 
\begin{equation}
\Delta    S_{\text{GED}}= -\int d^{d+1}x \,e\,  \frac{\mu^2}{2}  \left (  
h^{\mu\nu}A_{\mu}A_{\nu}-2\hat{v}^{\mu}A_{\mu}\varphi+2\tilde{\Phi}\varphi^{2}\right)\,.
\end{equation}
to the action \eqref{eq:GED2} of GED coupled to TNC. On flat TNC space this leads to the modified equations of motion
\begin{eqnarray}
\partial_{t}^{2}\varphi+\left(\partial_{i}\partial^{i}\tilde{\varphi}+\partial_{t}\partial_{i}a^{i}\right) & = & \mu^{2}\tilde{\varphi}\\
-\partial_{i}\partial^{i}\varphi & = & \mu^{2}\varphi\\
\partial_{t}\partial_{i}\varphi+\partial_{i}\partial_{k}a^{k}-\partial_{k}\partial^{k}a_{i} & = & -\mu^{2}a_{i}\,.
\end{eqnarray}
One could thus study how these terms affect the degrees of freedom and the symmetries of the theory.


A further natural direction would be to consider the
ultra-relativistic limit of electromagnetism, i.e. the Carrollian limit of Maxwell's theory (see e.g. \cite{Duval:2014uoa}).
The equations of motions follow again by appropriately scaling the fields, using Maxwell's equations \eqref{eq:Maxwell} and the limit
 $c \rightarrow 0$. 
In particular for the electric limit theory one takes $A_t=-\tilde\varphi$ and $A_i=ca_i$, leading to the equations of motion
\begin{equation}
\partial_i(\partial_t a_i+\partial_i\tilde\varphi)=0    \,,\qquad  \partial_t(\partial_t a_i+\partial_i\tilde\varphi)=0
\end{equation}
where we note that the first equation coincides with the first equation of the non-relativistic magnetic limit \eqref{eq:Mlimit}. 
This limit can also be taken at the level of the Maxwell action leading to an action proportional to 
$\frac{1}{2}(\partial_t a_i+\partial_i\tilde\varphi)^2$.  For the magnetic limit theory the fields scale as 
 $A_t=-\varphi$ and $ A_i=a_i/c$, leading to the equations of motion
 \begin{equation}
 \partial_t\partial_i a_i=0  \,,\qquad   \partial_t^2 a_i=0
\end{equation}
It would be interesting to study the symmetries of these theories  \cite{LŽvy1965}, their coupling to charged matter, and the covariant coupling to curved Carrollian geometry  (see e.g. \cite{Duval:2014uoa,Bekaert:2015xua,Hartong:2015xda}).

Finally, it would be interesting to apply our results to non-relativistic condensed matter systems, which could be relevant
in situations where the electromagnetic field is a static electric or magnetic field, so that there are no electromagnetic waves. 
In this context, Chern-Simons formulations of non-relativistic electrodynamics might be worthwhile to consider as well. 
It remains an intriguing open question whether GED is realized in concrete real-life systems.

\section*{Acknowledgements}

We want to thank Arjun Bagchi, Eric Bergshoeff, Dieter van Bleeken, Cynthia Keeler and Jan Rosseel for interesting discussions.
The work of GF is supported by the ERC STG grant 639220. 
The work of JH is partially supported by a Marina Solvay fellowship as well as by the advanced ERC grant `Symmetries and Dualities in Gravity and M-theory' of Marc Henneaux. 
The work of NO is supported in part by the Danish National Research Foundation project 
 ``New horizons in particle and condensed matter physics from black holes".

\addcontentsline{toc}{section}{\refname}

\begin{thebibliography}{10}

\bibitem{LeBellac:1973}
M.~L. Bellac and J.~M. L\'evy-Leblond, ``{Galilean Electromagnetism},'' {\em
  Nuovo Cim.} {\bf B14} (1973)  217--33.

\bibitem{Son:2013rqa}
D.~T. Son, ``{Newton-Cartan Geometry and the Quantum Hall Effect},''
\href{http://arxiv.org/abs/1306.0638}{{\tt arXiv:1306.0638
  [cond-mat.mes-hall]}}.

\bibitem{Christensen:2013lma}
M.~H. Christensen, J.~Hartong, N.~A. Obers, and B.~Rollier, ``{Torsional
  Newton-Cartan Geometry and Lifshitz Holography},''
  \href{http://dx.doi.org/10.1103/PhysRevD.89.061901}{{\em Phys.Rev.} {\bf D89}
  (2014)  061901},
\href{http://arxiv.org/abs/1311.4794}{{\tt arXiv:1311.4794 [hep-th]}}.

\bibitem{Christensen:2013rfa}
M.~H. Christensen, J.~Hartong, N.~A. Obers, and B.~Rollier, ``{Boundary
  Stress-Energy Tensor and Newton-Cartan Geometry in Lifshitz Holography},''
  \href{http://dx.doi.org/10.1007/JHEP01(2014)057}{{\em JHEP} {\bf 1401} (2014)
   057},
\href{http://arxiv.org/abs/1311.6471}{{\tt arXiv:1311.6471 [hep-th]}}.

\bibitem{Hartong:2014oma}
J.~Hartong, E.~Kiritsis, and N.~A. Obers, ``{Lifshitz spaceÐtimes for
  Schršdinger holography},''
  \href{http://dx.doi.org/10.1016/j.physletb.2015.05.010}{{\em Phys. Lett.}
  {\bf B746} (2015)  318--324},
\href{http://arxiv.org/abs/1409.1519}{{\tt arXiv:1409.1519 [hep-th]}}.

\bibitem{Geracie:2014nka}
M.~Geracie, D.~T. Son, C.~Wu, and S.-F. Wu, ``{Spacetime Symmetries of the
  Quantum Hall Effect},''
  \href{http://dx.doi.org/10.1103/PhysRevD.91.045030}{{\em Phys.Rev.} {\bf D91}
  (2015)  045030},
\href{http://arxiv.org/abs/1407.1252}{{\tt arXiv:1407.1252
  [cond-mat.mes-hall]}}.

\bibitem{Jensen:2014aia}
K.~Jensen, ``{On the coupling of Galilean-invariant field theories to curved
  spacetime},''
\href{http://arxiv.org/abs/1408.6855}{{\tt arXiv:1408.6855 [hep-th]}}.

\bibitem{Hartong:2014pma}
J.~Hartong, E.~Kiritsis, and N.~A. Obers, ``{Schršdinger Invariance from
  Lifshitz Isometries in Holography and Field Theory},''
  \href{http://dx.doi.org/10.1103/PhysRevD.92.066003}{{\em Phys. Rev.} {\bf
  D92} (2015)  066003},
\href{http://arxiv.org/abs/1409.1522}{{\tt arXiv:1409.1522 [hep-th]}}.

\bibitem{Hartong:2015wxa}
J.~Hartong, E.~Kiritsis, and N.~A. Obers, ``{Field Theory on Newton-Cartan
  Backgrounds and Symmetries of the Lifshitz Vacuum},''
  \href{http://dx.doi.org/10.1007/JHEP08(2015)006}{{\em JHEP} {\bf 08} (2015)
  006},
\href{http://arxiv.org/abs/1502.00228}{{\tt arXiv:1502.00228 [hep-th]}}.

\bibitem{Duval:2014uoa}
C.~Duval, G.~Gibbons, P.~Horvathy, and P.~Zhang, ``{Carroll versus Newton and
  Galilei: two dual non-Einsteinian concepts of time},''
  \href{http://dx.doi.org/10.1088/0264-9381/31/8/085016}{{\em
  Class.Quant.Grav.} {\bf 31} (2014)  085016},
\href{http://arxiv.org/abs/1402.0657}{{\tt arXiv:1402.0657 [gr-qc]}}.

\bibitem{Geracie:2015dea}
M.~Geracie, K.~Prabhu, and M.~M. Roberts, ``{Curved non-relativistic
  spacetimes, Newtonian gravitation and massive matter},''
  \href{http://dx.doi.org/10.1063/1.4932967}{{\em J. Math. Phys.} {\bf 56}
  (2015) no.~10, 103505},
\href{http://arxiv.org/abs/1503.02682}{{\tt arXiv:1503.02682 [hep-th]}}.

\bibitem{Bleeken:2015ykr}
D.~Van~den Bleeken and C.~Yunus, ``{Newton-Cartan, Galileo-Maxwell and
  Kaluza-Klein},'' \href{http://dx.doi.org/10.1088/0264-9381/33/13/137002}{{\em
  Class. Quant. Grav.} {\bf 33} (2016) no.~13, 137002},
\href{http://arxiv.org/abs/1512.03799}{{\tt arXiv:1512.03799 [hep-th]}}.

\bibitem{Bergshoeff:2015sic}
E.~Bergshoeff, J.~Rosseel, and T.~Zojer, ``{Non-relativistic fields from
  arbitrary contracting backgrounds},''
  \href{http://dx.doi.org/10.1088/0264-9381/33/17/175010}{{\em Class. Quant.
  Grav.} {\bf 33} (2016) no.~17, 175010},
\href{http://arxiv.org/abs/1512.06064}{{\tt arXiv:1512.06064 [hep-th]}}.

\bibitem{Festuccia:2016awg}
G.~Festuccia, D.~Hansen, J.~Hartong, and N.~A. Obers, ``{Torsional
  Newton-Cartan Geometry from the Noether Procedure},''
\href{http://arxiv.org/abs/1607.01926}{{\tt arXiv:1607.01926 [hep-th]}}.

\bibitem{Taylor:2015glc}
M.~Taylor, ``{Lifshitz holography},'' {\em Class. Quant. Grav.} {\bf 33} (2016)
  no.~3, 033001,
\href{http://arxiv.org/abs/1512.03554}{{\tt arXiv:1512.03554 [hep-th]}}.

\bibitem{Hartong:2015zia}
J.~Hartong and N.~A. Obers, ``{Ho{\v r}ava-Lifshitz gravity from dynamical
  Newton-Cartan geometry},''
  \href{http://dx.doi.org/10.1007/JHEP07(2015)155}{{\em JHEP} {\bf 07} (2015)
  155},
\href{http://arxiv.org/abs/1504.07461}{{\tt arXiv:1504.07461 [hep-th]}}.

\bibitem{Griffin:2012qx}
T.~Griffin, P.~Horava, and C.~M. Melby-Thompson, ``{Lifshitz Gravity for
  Lifshitz Holography},''
  \href{http://dx.doi.org/10.1103/PhysRevLett.110.081602}{{\em Phys.Rev.Lett.}
  {\bf 110} (2013) no.~8, 081602},
\href{http://arxiv.org/abs/1211.4872}{{\tt arXiv:1211.4872 [hep-th]}}.

\bibitem{Janiszewski:2012nf}
S.~Janiszewski and A.~Karch, ``{String Theory Embeddings of Nonrelativistic
  Field Theories and Their Holographic Horava Gravity Duals},''
  \href{http://dx.doi.org/10.1103/PhysRevLett.110.081601}{{\em Phys.Rev.Lett.}
  {\bf 110} (2013) no.~8, 081601},
\href{http://arxiv.org/abs/1211.0010}{{\tt arXiv:1211.0010 [hep-th]}}.

\bibitem{Hartong:2016yrf}
J.~Hartong, Y.~Lei, and N.~A. Obers, ``{Non-Relativistic Chern-Simons Theories
  and Three-Dimensional Horava-Lifshitz Gravity},''
\href{http://arxiv.org/abs/1604.08054}{{\tt arXiv:1604.08054 [hep-th]}}. To appear in PRD. 

\bibitem{Andringa:2013mma}
R.~Andringa, E.~A. Bergshoeff, J.~Rosseel, and E.~Sezgin, ``{3D NewtonÐCartan
  supergravity},'' \href{http://dx.doi.org/10.1088/0264-9381/30/20/205005}{{\em
  Class. Quant. Grav.} {\bf 30} (2013)  205005},
\href{http://arxiv.org/abs/1305.6737}{{\tt arXiv:1305.6737 [hep-th]}}.

\bibitem{Bergshoeff:2015uaa}
E.~Bergshoeff, J.~Rosseel, and T.~Zojer, ``{NewtonÐCartan (super)gravity as a
  non-relativistic limit},''
  \href{http://dx.doi.org/10.1088/0264-9381/32/20/205003}{{\em Class. Quant.
  Grav.} {\bf 32} (2015) no.~20, 205003},
\href{http://arxiv.org/abs/1505.02095}{{\tt arXiv:1505.02095 [hep-th]}}.

\bibitem{Bergshoeff:2015ija}
E.~Bergshoeff, J.~Rosseel, and T.~Zojer, ``{Newton-Cartan supergravity with
  torsion and Schršdinger supergravity},''
  \href{http://dx.doi.org/10.1007/JHEP11(2015)180}{{\em JHEP} {\bf 11} (2015)
  180},
\href{http://arxiv.org/abs/1509.04527}{{\tt arXiv:1509.04527 [hep-th]}}.

\bibitem{Knodel:2015byb}
G.~Knodel, P.~Lisbao, and J.~T. Liu, ``{Rigid Supersymmetric Backgrounds of
  3-dimensional Newton-Cartan Supergravity},''
  \href{http://dx.doi.org/10.1007/JHEP06(2016)028}{{\em JHEP} {\bf 06} (2016)
  028},
\href{http://arxiv.org/abs/1512.04961}{{\tt arXiv:1512.04961 [hep-th]}}.

\bibitem{Bergshoeff:2016lwr}
E.~A. Bergshoeff and J.~Rosseel, ``{Three-Dimensional Extended Bargmann
  Supergravity},'' \href{http://dx.doi.org/10.1103/PhysRevLett.116.251601}{{\em
  Phys. Rev. Lett.} {\bf 116} (2016) no.~25, 251601},
\href{http://arxiv.org/abs/1604.08042}{{\tt arXiv:1604.08042 [hep-th]}}.

\bibitem{Festuccia:2011ws}
G.~Festuccia and N.~Seiberg, ``{Rigid Supersymmetric Theories in Curved
  Superspace},'' \href{http://dx.doi.org/10.1007/JHEP06(2011)114}{{\em JHEP}
  {\bf 06} (2011)  114},
\href{http://arxiv.org/abs/1105.0689}{{\tt arXiv:1105.0689 [hep-th]}}.

\bibitem{Pestun:2007rz}
V.~Pestun, ``{Localization of gauge theory on a four-sphere and supersymmetric
  Wilson loops},'' \href{http://dx.doi.org/10.1007/s00220-012-1485-0}{{\em
  Commun. Math. Phys.} {\bf 313} (2012)  71--129},
\href{http://arxiv.org/abs/0712.2824}{{\tt arXiv:0712.2824 [hep-th]}}.

\bibitem{Santos:2004pq}
E.~S. Santos, M.~de~Montigny, F.~C. Khanna, and A.~E. Santana, ``{Galilean
  covariant Lagrangian models},''
\href{http://dx.doi.org/10.1088/0305-4470/37/41/011}{{\em J. Phys.} {\bf A37}
  (2004)  9771--9789}.

\bibitem{DeMontigny:2005oib}
M.~De~Montigny and G.~Rousseaux, ``{On the electrodynamics of moving bodies at
  low velocities},'' \href{http://dx.doi.org/10.1088/0143-0807/27/4/007}{{\em
  Eur. J. Phys.} {\bf 27} (2006)  755--768},
\href{http://arxiv.org/abs/physics/0512200}{{\tt arXiv:physics/0512200
  [physics]}}.

\bibitem{Rousseaux:2013}
G.~Rousseaux, ``{Forty years of Galilean Electromagnetism (1973Ð2013)},''
\href{http://dx.doi.org/10.1140/epjp/i2013-13081-5}{{\em Eur. Phys. J. Plus}
  {\bf 128:81} (2013)  }.

\bibitem{Jensen:2014wha}
K.~Jensen and A.~Karch, ``{Revisiting non-relativistic limits},''
  \href{http://dx.doi.org/10.1007/JHEP04(2015)155}{{\em JHEP} {\bf 04} (2015)
  155},
\href{http://arxiv.org/abs/1412.2738}{{\tt arXiv:1412.2738 [hep-th]}}.

\bibitem{Duval:1984cj}
C.~Duval, G.~Burdet, H.~Kunzle, and M.~Perrin, ``{Bargmann Structures and
  Newton-Cartan Theory},''
\href{http://dx.doi.org/10.1103/PhysRevD.31.1841}{{\em Phys.Rev.} {\bf D31}
  (1985)  1841}.

\bibitem{Julia:1994bs}
B.~Julia and H.~Nicolai, ``{Null Killing vector dimensional reduction and
  Galilean geometrodynamics},''
  \href{http://dx.doi.org/10.1016/0550-3213(94)00584-2}{{\em Nucl.Phys.} {\bf
  B439} (1995)  291--326},
\href{http://arxiv.org/abs/hep-th/9412002}{{\tt arXiv:hep-th/9412002
  [hep-th]}}.

\bibitem{Duval:1990hj}
C.~Duval, G.~W. Gibbons, and P.~Horvathy, ``{Celestial mechanics, conformal
  structures and gravitational waves},''
  \href{http://dx.doi.org/10.1103/PhysRevD.43.3907}{{\em Phys.Rev.} {\bf D43}
  (1991)  3907--3922},
\href{http://arxiv.org/abs/hep-th/0512188}{{\tt arXiv:hep-th/0512188
  [hep-th]}}.

\bibitem{Bagchi:2014ysa}
A.~Bagchi, R.~Basu, and A.~Mehra, ``{Galilean Conformal Electrodynamics},''
  \href{http://dx.doi.org/10.1007/JHEP11(2014)061}{{\em JHEP} {\bf 11} (2014)
  061},
\href{http://arxiv.org/abs/1408.0810}{{\tt arXiv:1408.0810 [hep-th]}}.

\bibitem{Duval:2009vt}
C.~Duval and P.~A. Horvathy, ``{Non-relativistic conformal symmetries and
  Newton-Cartan structures},''
  \href{http://dx.doi.org/10.1088/1751-8113/42/46/465206}{{\em J.Phys.} {\bf
  A42} (2009)  465206},
\href{http://arxiv.org/abs/0904.0531}{{\tt arXiv:0904.0531 [math-ph]}}.

\bibitem{jackson_classical_1999}
J.~D. Jackson, {\em Classical electrodynamics}.
\newblock Wiley, New York, {NY}, 3rd ed.~ed., 1999.
\newblock \url{http://cdsweb.cern.ch/record/490457}.

\bibitem{Brown1999}
H.~R. Brown and P.~R. Holland, ``The galilean covariance of quantum mechanics
  in the case of external fields,''
  \href{http://dx.doi.org/http://dx.doi.org/10.1119/1.19227}{{\em American
  Journal of Physics} {\bf 67} (1999) no.~3, 204--214}.
  \url{http://scitation.aip.org/content/aapt/journal/ajp/67/3/10.1119/1.19227}.

\bibitem{Goldin2001321}
G.~A. Goldin and V.~M. Shtelen, ``On galilean invariance and nonlinearity in
  electrodynamics and quantum mechanics,''
  \href{http://dx.doi.org/http://dx.doi.org/10.1016/S0375-9601(01)00017-2}{{\em
  Physics Letters A} {\bf 279} (2001) no.~5–6, 321 -- 326}.
  \url{http://www.sciencedirect.com/science/article/pii/S0375960101000172}.

\bibitem{Bagchi:2009my}
A.~Bagchi and R.~Gopakumar, ``{Galilean Conformal Algebras and AdS/CFT},''
  \href{http://dx.doi.org/10.1088/1126-6708/2009/07/037}{{\em JHEP} {\bf 0907}
  (2009)  037},
\href{http://arxiv.org/abs/0902.1385}{{\tt arXiv:0902.1385 [hep-th]}}.

\bibitem{Martelli:2009uc}
D.~Martelli and Y.~Tachikawa, ``{Comments on Galilean conformal field theories
  and their geometric realization},''
  \href{http://dx.doi.org/10.1007/JHEP05(2010)091}{{\em JHEP} {\bf 05} (2010)
  091},
\href{http://arxiv.org/abs/0903.5184}{{\tt arXiv:0903.5184 [hep-th]}}.

\bibitem{Hofman:2011zj}
D.~M. Hofman and A.~Strominger, ``{Chiral Scale and Conformal Invariance in 2D
  Quantum Field Theory},''
  \href{http://dx.doi.org/10.1103/PhysRevLett.107.161601}{{\em Phys.Rev.Lett.}
  {\bf 107} (2011)  161601},
\href{http://arxiv.org/abs/1107.2917}{{\tt arXiv:1107.2917 [hep-th]}}.

\bibitem{Detournay:2012pc}
S.~Detournay, T.~Hartman, and D.~M. Hofman, ``{Warped Conformal Field
  Theory},'' \href{http://dx.doi.org/10.1103/PhysRevD.86.124018}{{\em
  Phys.Rev.} {\bf D86} (2012)  124018},
\href{http://arxiv.org/abs/1210.0539}{{\tt arXiv:1210.0539 [hep-th]}}.

\bibitem{Bergshoeff:2014uea}
E.~A. Bergshoeff, J.~Hartong, and J.~Rosseel, ``{Torsional NewtonÐCartan
  geometry and the Schršdinger algebra},''
  \href{http://dx.doi.org/10.1088/0264-9381/32/13/135017}{{\em Class. Quant.
  Grav.} {\bf 32} (2015) no.~13, 135017},
\href{http://arxiv.org/abs/1409.5555}{{\tt arXiv:1409.5555 [hep-th]}}.

\bibitem{Bekaert:2014bwa}
X.~Bekaert and K.~Morand, ``{Connections and dynamical trajectories in
  generalised Newton-Cartan gravity I. An intrinsic view},''
  \href{http://dx.doi.org/10.1063/1.4937445}{{\em J. Math. Phys.} {\bf 57}
  (2016) no.~2, 022507},
\href{http://arxiv.org/abs/1412.8212}{{\tt arXiv:1412.8212 [hep-th]}}.

\bibitem{Kuchar:1980tw}
K.~Kuchar, ``{Gravitation, Geometry and Nonrelativistic Quantum Theory},''
\href{http://dx.doi.org/10.1103/PhysRevD.22.1285}{{\em Phys.Rev.} {\bf D22}
  (1980)  1285--1299}.

\bibitem{Bergshoeff:2014gja}
E.~Bergshoeff, J.~Gomis, M.~Kovacevic, L.~Parra, J.~Rosseel, {\em et al.},
  ``{Nonrelativistic superparticle in a curved background},''
  \href{http://dx.doi.org/10.1103/PhysRevD.90.065006}{{\em Phys.Rev.} {\bf D90}
  (2014) no.~6, 065006},
\href{http://arxiv.org/abs/1406.7286}{{\tt arXiv:1406.7286 [hep-th]}}.

\bibitem{Fuini:2015yva}
J.~F. Fuini, A.~Karch, and C.~F. Uhlemann, ``{Spinor fields in general
  Newton-Cartan backgrounds},''
  \href{http://dx.doi.org/10.1103/PhysRevD.92.125036}{{\em Phys. Rev.} {\bf
  D92} (2015) no.~12, 125036},
\href{http://arxiv.org/abs/1510.03852}{{\tt arXiv:1510.03852 [hep-th]}}.

\bibitem{Bagchi:2015qcw}
A.~Bagchi, R.~Basu, A.~Kakkar, and A.~Mehra, ``{Galilean Yang-Mills Theory},''
  \href{http://dx.doi.org/10.1007/JHEP04(2016)051}{{\em JHEP} {\bf 04} (2016)
  051},
\href{http://arxiv.org/abs/1512.08375}{{\tt arXiv:1512.08375 [hep-th]}}.

\bibitem{LŽvy1965}
J.-M. Levy-Leblond, ``Une nouvelle limite non-relativiste du groupe de
  poincar\'e,'' {\em Annales de l'institut Henri Poincar\'e (A) Physique
  th\'eorique} {\bf 3} (1965) no.~1, 1--12. \url{http://eudml.org/doc/75509}.

\bibitem{Bekaert:2015xua}
X.~Bekaert and K.~Morand, ``{Connections and dynamical trajectories in
  generalised Newton-Cartan gravity II. An ambient perspective},''
\href{http://arxiv.org/abs/1505.03739}{{\tt arXiv:1505.03739 [hep-th]}}.

\bibitem{Hartong:2015xda}
J.~Hartong, ``{Gauging the Carroll Algebra and Ultra-Relativistic Gravity},''
  \href{http://dx.doi.org/10.1007/JHEP08(2015)069}{{\em JHEP} {\bf 08} (2015)
  069},
\href{http://arxiv.org/abs/1505.05011}{{\tt arXiv:1505.05011 [hep-th]}}.

\end{thebibliography}

\providecommand{\href}[2]{#2}\begingroup\raggedright\endgroup

\end{document}